\documentclass[aps,prb,twocolumn,groupedaddress,showpacs,amsmath,amssymb]{revtex4}

\usepackage[english]{babel}

\usepackage{graphicx}
\usepackage{hyperref}
\usepackage{subfigure} 
\usepackage{multirow} 

\DeclareMathOperator{\Var}{Var}
\DeclareMathOperator{\Cov}{Cov}

\begin{document}

\raggedbottom

\title{Lattice charge models and core level shifts in disordered alloys}
\author{T. L. Underwood}
\author{R. J. Cole}
        
\affiliation{School of Physics and Astronomy, SUPA, The University of Edinburgh, Edinburgh,
EH9 3JZ, UK }

\date{\today}


\begin{abstract}
Differences in core level binding energies between atoms belonging to the same chemical species can be related to differences in their intra- and extra-atomic charge
distributions, and differences in how their core holes are screened. With this in mind, we consider the charge-excess functional model (CEFM) for net atomic 
charges in alloys [E. Bruno \emph{et al.}, Phys. Rev. Lett. \textbf{91}, 166401 (2003)].
We begin by deriving the CEFM energy function in order to elucidate the approximations which underpin this model.
We thereafter consider the particular case of the CEFM in which the strength of the `local interactions' within all atoms are the same.
We show that for binary alloys the ground state charges of this model can be expressed in terms of charge transfer between all pairs of \emph{unlike}
atoms analogously to the linear charge model [R. Magri \emph{et al.}, Phys. Rev. B \textbf{42}, 11388 (1990)]. Hence the model considered
is a generalization of the linear charge model for alloys containing more than two chemical species. 
We then determine the model's unknown `geometric factors' over a wide range of parameter space. These quantities are linked to the nature of 
charge screening in the model, and we illustrate that the screening becomes increasingly universal as the strength of the local interactions
is increased.
We then use the model to derive analytical expressions for various physical quantities, including the Madelung energy and the disorder broadening
in the core level binding energies. These expressions are applied to ternary random alloys, for which it is shown that the Madelung 
energy and magnitude of disorder broadening are maximized at the composition at which the two species with the largest `electronegativity difference'
are equal, while the remaining species having a vanishing concentration. This result is somewhat counterintuitive with regards to the disorder 
broadening since it does not correspond to the composition with the highest entropy.
Finally, the model is applied to CuPd and CuZn random alloys. The model is used to deduce the effective radii associated with valence electron charge 
transfer for Cu, Pd and Zn in these systems for use in the ESCA potential model of X-ray photoelectron spectroscopy. The effective radii are found to 
be $R_1/3$, where $R_1$ is the nearest neighbor distance, with only small variations between chemical elements and between different systems. The model
provides a framework for rationalising the disorder broadenings in these systems: they can be understood in terms of an 
interplay between the broadening in the Madelung potentials and the broadening in the intra-atomic electrostatic potentials.
\end{abstract}

\maketitle

\section{Introduction}\label{sec:intro}
Disordered alloys, in which the nuclei form an approximate crystal lattice, but in which the pattern formed by considering the chemical species
of the nuclei is not periodic, are of fundamental importance to metallurgy and nanotechnology. It is well known that the binding energies of core 
levels depend on the environment of the nucleus to which they are bound. Therefore X-ray photoelectron spectroscopy (XPS), which provides the 
distribution of core level binding energies within a sample, can in theory be used as a probe of specific environments in disordered alloys. 
The increased resolution and bulk sensitivity afforded by the latest instrumentation and synchrotron light sources have made such
`environment-resolved spectroscopy' an exciting prospect \cite{Holmstrom_2006,Granroth_2009,Olovsson_2011,Granroth_2011}. 
However, accurate interpretation of experimental core level spectra requires a solid understanding of the relationship between a site's%
\footnote{In this paper we only consider systems in which the nuclei form a perfect crystal lattice, and we use the term \emph{site} to refer to the 
collective contents of the Wigner-Seitz cell centered on a particular nucleus.}
core level binding energies, its local electronic structure, and its environment. 

Traditionally, mean field approximations such as the single site coherent potential approximation (SSCPA) \cite{Soven_1967,Johnson_1986,Johnson_1990}
have been used in conjunction with density functional theory (DFT) \cite{Hohenberg_1964,Kohn_1965} to model disordered alloys. While these
approximations can be used to determine the average properties over groups of sites belonging to each of an alloy's constituent species, they
cannot provide details of the \emph{distribution} of site properties about these averages. It was not until the 1990s that developments in computer
hardware and so-called order-$N$ methods, such as the locally self-consistent multiple-scattering (LSMS) \cite{Wang_1995} and the locally 
self-consistent Green's function (LSGF) \cite{Abrikosov_1996,Abrikosov_1997} methods, allowed the distribution of site properties
within disordered alloys to be determined in an \emph{ab initio} fashion. These calculations revealed a surprisingly simple relationship
between the net charge on a site and its Madelung potential - the \emph{Q-V} relations \cite{Faulkner_1995,Faulkner_1997}: 
\begin{equation}\label{QV_rel}
V_i=-a_iQ_i+k_i,
\end{equation}
where $V_i$ is the Madelung potential of site $i$, $Q_i$ is the net charge on site $i$, and $a_i$ and $k_i$ take the values $a_X$ and $k_X$
respectively if site $i$ belongs to species $X$. The \emph{Q-V} relations were found to hold over a wide range of disordered alloys, with
various combinations of constituent species, concentrations of each species and degrees of substitutional disorder. Furthermore,
the values of $a_X$ and $k_X$ for each species are seemingly the same for all alloys, even ordered, which have the same species concentrations
and underlying crystal lattice \cite{Faulkner_1995}. More recently, Ruban and Skriver (RS) \cite{Ruban_2002} discovered that there is a universality
to the \emph{Q-V} relations: for $X$ sites in any alloy, the values of $a_X$ and $k_X$ obey
\footnote{Throughout this work we use Hartree atomic units unless otherwise stated.}
\begin{equation}\label{us_aR}
a_XR_{\text{WS}}\approx 1.6
\end{equation}
and
\begin{equation}\label{us_kR}
k_XR_{\text{WS}}/Q^{\text{SSCPA}}_X\approx 1.6
\end{equation}
respectively, where $Q^{\text{SSCPA}}_X$ is the charge of an $X$ site obtained from a conventional SSCPA calculation for the alloy in question, and
$R_{\text{WS}}$ is the alloy's Wigner-Seitz radius. This result was obtained using the single site LSGF (SSLSGF) method - essentially a generalization
of the SSCPA in which each site is treated as a separate species for the purposes of constructing the mean-field effective medium and evaluating each site's
properties. RS attributed the above result to a universal mechanism of screening, which we will now describe. Consider what happens if we perturb
$Q_i$ by an amount $\delta Q_0$, and allow the charge on all other sites to relax so to minimize the total energy of the system.
Firstly, RS discovered that the induced change in charge for all sites in the $\beta$th nearest neighbor shell of $i$
\footnote{The nearest neighbor shell of site $i$ is the set of sites, not including $i$ itself, which are closest to $i$. The $\beta$th
nearest neighbor shell is the set of sites, not including $i$ or those sites in shells $(\beta-1),(\beta-2),\dotsc,2,1$ of $i$, which are
closest to $i$.}
are the same, which we denote as $\delta Q_{\beta}$. RS also discovered that, regardless of the choice of site and the particular alloy,
\begin{equation}\label{us_function}
\delta Q_{\beta}/\delta Q_0=u(R_{\beta}/R_{\text{WS}}),
\end{equation}
where $R_{\beta}$ is the distance from site $i$ to its $\beta$th nearest neighbor shell, and $u$ is a universal function. In other words,
the values of $\delta Q_{\beta}/\delta Q_0$ obtained from considering all sites in all alloys, when plotted against $R_{\beta}/R_{\text{WS}}$, lie
upon a single curve.

The existence of the \emph{Q-V} relations and universal screening hints that the charge distribution and screening in disordered alloys could be
accurately described by a simple model. Differences in core level binding energies between sites belonging to the same chemical species can be 
related, via the `ESCA potential model' \cite{siegbahn,Fadley_1968,Gelius_1974,Williams_1978}, to differences in their intra- and extra-atomic charge 
distributions, and differences in how their core holes are screened. Therefore the prospect exists of developing a simple model to describe the 
distribution of core level shifts%
\footnote{In this paper we consider only metal-alloy core level shifts, which for a core level bound to an $X$ site $i$ is defined as
$\Delta E^{\text{B}}_i=E^{\text{B}}_i-E^{\text{B}}_{\text{metal}}$, where $E^{\text{B}}_i$ is the binding energy of the core level, and $E^{\text{B}}_{\text{metal}}$
is the binding energy of the corresponding core level in an $X$ metal.}
in disordered alloys. It is worth pursuing such simple models since they can act as a faster alternative to computationally expensive
\emph{ab initio} methods in certain situations, aid in the interpretation of experimental results, and illuminate the relevant underlying physics. 

Before the discovery of the \emph{Q-V} relations, the linear charge model (LCM) \cite{Magri_1990} was proposed for calculating the values of $Q_i$ in
binary alloys. This model is based on the assumption that \emph{unlike} nearest neighbors transfer a fixed quantity of charge. The results of order-$N$
calculations have been used both to laud and to criticize the LCM \cite{Faulkner_1997,Wolverton_1996}; in terms of quantitative predictions, it has 
been shown that, at best, the LCM is only in semiquantitative agreement with the results of order-$N$ calculations \cite{Wolverton_1996,Underwood_2009}. 
However, despite the shortcomings of the LCM, it certainly performs far better than might be expected from its simplicity. With this in mind, several 
generalizations of the LCM have been proposed \cite{Wolverton_1996,Underwood_2009}. The most successful of these is the multi-shell linear charge model 
(MLCM) \cite{Wolverton_1996,Underwood_2009}, in which each pair of unlike sites separated by $R_{\beta}$ transfer an amount of charge $2\lambda_{\beta}$,
with the convention that the $A$ site loses $2\lambda_{\beta}$ while the $B$ site gains $2\lambda_{\beta}$. Assuming that no charge transfer occurs 
between pairs of sites separated by more than $R_{\beta_{\text{max}}}$, then the free parameters in the model become 
$\lambda_1,\lambda_2,\dotsc,\lambda_{\beta_{\text{max}}}$, and the following expression for $Q_i$ applies:
\begin{equation}\label{Q_MLCM}
Q_i=2S_i\sum_{\beta=1}^{\beta_{\text{max}}}\lambda_{\beta}N_{i\beta},
\end{equation}
where $N_{i\beta}$ is the number of unlike sites in the $\beta$th nearest neighbor shell of site $i$, and $S_i=-1$ if $i$ belongs to species $A$ and +1
if $i$ belongs to species $B$. The MLCM has been shown to reproduce the values of $Q_i$ and $V_i$ obtained from order-$N$ calculations to a high
degree of accuracy - when the model is appropriately parameterized \cite{Wolverton_1996}. However, the MLCM has been criticized for the fact that
it has too many free parameters \cite{Faulkner_1997,Bruno_2004}. In an attempt to reduce the number of free parameters in the MLCM, it was proposed in 
Ref. \onlinecite{Underwood_2009} that they be constrained such that \emph{Q-V} relations are obeyed `as closely as is possible' for
the chosen value of $\beta_{\text{max}}$. We will refer to this particular case of the MLCM as the optimized linear charge model (OLCM). In terms of
quantitative predictions, the OLCM performs well: calibrated appropriately, it gives good quantitative agreement with LSMS results 
\cite{Underwood_2009}. Furthermore, there is seemingly a universal mechanism to the screening in the OLCM which is in semiquantitative agreement with 
that described by RS \cite{Underwood_2009}.

An alternative approach to the LCM and its derivatives can be found in the charge-excess functional model (CEFM) \cite{Bruno_2003}. In the CEFM, the
values of $Q_i$ are postulated to be those which minimize the energy function
\begin{equation}\label{E_gen_def}
E=E_L+E_M,
\end{equation}
where
\begin{equation}\label{EL_gen_def}
E_L=\frac{1}{2}\sum_ia_i(Q_i-b_i)^2
\end{equation}
is known as the local energy,
\begin{equation}\label{EM_gen_def}
E_M=\frac{1}{2}\sum_iM_{ij}Q_iQ_j
\end{equation}
is the Madelung energy,
\begin{equation}\label{M_def}
M_{ij}=
\begin{cases}
0 & \text{if $j=i$}, \\
1/|\mathbf{R}_i-\mathbf{R}_j| & \text{otherwise}
\end{cases}
\end{equation}
is the Madelung matrix, $\mathbf{R}_i$ is the position of nucleus $i$, and $a_i$ is the strength of the `local interactions' within site $i$ which
act to keep the charge of site $i$ at its `bare' value $b_i$. The values of $a_i$ and $b_i$ for all sites belonging to the same species $X$ are
required to take the same values $a_X$ and $b_X$ respectively. Minimizing $E$ subject to the constraint of charge neutrality, i.e.
\begin{equation}\label{sumQi=0}
\sum_iQ_i=0,
\end{equation}
leads to Eqn. \eqref{QV_rel} with
\begin{equation}\label{ki_def}
k_i=a_ib_i+\mu,
\end{equation}
where $\mu$ is a Lagrange multiplier added to ensure that Eqn. \eqref{sumQi=0} is obeyed. Hence the \emph{Q-V} relations are implicit in the CEFM: $V_i$
and $Q_i$ for $X$ sites will \emph{always} form a \emph{Q-V} relation with gradient $-a_X$ and intercept $k_X=a_Xb_X+\mu$. It is therefore perhaps 
unsurprising that, if the values of $a_X$ and $b_X$ for each species - which are the free parameters in the CEFM - are derived from order-$N$ 
calculations, then the CEFM gives an extremely accurate description of disordered alloys \cite{Bruno_2003,Drchal_2006,Bruno_2008}. However,
the strength of the CEFM comes from the fact that, as mentioned earlier, the free parameters are seemingly transferable between systems with the same 
species concentrations and underlying crystal lattice. Hence the CEFM is an efficient method for evaluating differences in the charge distribution
\cite{Bruno_2003,Drchal_2006} and energies \cite{Bruno_2008} between such systems, with potential applications including Monte Carlo simulations
\cite{Bruno_2009}.

The layout of this paper is as follows. In Sec. \ref{sec:underlying_approx} we provide a derivation of the CEFM energy functional which elucidates
the underlying assumptions of the model. The rest of the paper concerns itself with the particular case of the CEFM in which $a_i=a$ for all $i$. In
Sec. \ref{sec:NRA-CEFM} we derive some fundamental properties of the model which will be used throughout this paper, and discuss the model's accuracy
and its relationship to the OLCM. In Sec. \ref{sec:screening} we determine the model's unknown `geometric factors' and examine the nature of the 
screening in the model. In Sec. \ref{sec:analytical} analytical expressions are derived for various physical quantities, including: the variance in 
$Q_i$ for $X$ sites; the Madelung energy; and the magnitude of the $X$ initial state core level `disorder broadening', i.e. the width of the initial
state core level binding energy distribution associated with $X$ sites.
Furthermore, some of these expressions are used to investigate how the aforementioned physical quantities vary with the concentrations of the three
constituent species in a ternary random alloy.%
\footnote{By a `random alloy' we are referring to the special case of a disordered alloy in which there are no correlations at all between the species 
occupying each site.}
In Sec. \ref{sec:ab_initio} we apply the model to CuZn and CuPd random alloys in order to deduce the effective radii associated with 
valence electron charge transfer for each species in these systems, for use in the ESCA potential model. We also use these results to add insight into
the disorder broadening phenomenon more generally. Finally, in Sec. \ref{sec:summary} we give a summary of our key findings.

\section{Underlying approximations of the CEFM}\label{sec:underlying_approx}
The CEFM energy function of Eqns. \eqref{E_gen_def}, \eqref{EL_gen_def} and \eqref{EM_gen_def} has been derived from the atomic perspective in Ref. 
\onlinecite{Underwood_2009}. It has also been derived in Ref. \onlinecite{Bruno_2008} within the framework of DFT and multiple
scattering theory using a mean field approach. Here we present a complementary derivation of the energy function which encompasses both
perspectives, and elucidates the approximations which underpin the CEFM.
Consider a system of nuclei which form an infinite \emph{undistorted} crystal lattice. Taking site $i$ to be the Wigner-Seitz cell centered
on nucleus $i$, let
\begin{equation}
L_i\equiv\int_i d\mathbf{r}\;n(\mathbf{r})
\end{equation}
denote the total number of electrons within site $i$, where the `$i$' subscript on the integral signifies that it is over all positions
within site $i$, and $n(\mathbf{r})$ is the electron density at position $\mathbf{r}$. Note that 
\begin{equation}
Q_i=z_i-L_i,
\end{equation}
where $z_i$ is the atomic number of nucleus $i$. Making the assumption that the electron density within each site is spherically symmetric about the 
site's nucleus, which we will refer to as the \emph{spherical approximation}, the electron density within site $i$ can be characterized by $L_i$ and 
some function $s_i(r)$ which describes the \emph{radial distribution} of electrons within the site. Specifically, $L_is_i(r)$ is the electron density
at distance $r$ from $\mathbf{R}_i$, where $s_i(r)$ is constrained to obey
\begin{equation}\label{varsigma_const}
\int_0d\mathbf{r}\;s_i(|\mathbf{r}|)=1
\end{equation}
such that the total electron density within site $i$ integrates to $L_i$, where we have chosen site 0 to have $\mathbf{R}_0=\mathbf{0}$. Let us now 
\emph{define} $E_L$ to be the contribution to the total energy $E$ other than the Madelung energy: $E$ consists of the electronic kinetic energy, 
the intra-site Coulomb energy and the electronic exchange-correlation energy. With the above in mind, consider the contribution to $E_L$ from site $i$,
which we will denote as $E_{L,i}$. The intra-site Coulomb energy associated with site $i$ depends only on $z_i$, $L_i$ and $s_i(r)$; and the electronic 
kinetic and exchange-correlation energies associated with the electron density within site $i$ depend on the electron density throughout the 
entire system - which is completely determined by the underlying lattice and the quantities $L_j$ and $s_j(r)$ for all $j$. We will henceforth
consider the underlying lattice to be fixed, i.e. we will not treat it as a free parameter. In this case $E_{L,i}$ is a lattice-dependent functional
of the quantities $z_i$, and $L_j$ and $s_j(r)$ for all $j$. We will now assume that $E_{L,i}$ is a \emph{system-dependent} functional $\mathcal{E}$ 
\emph{only} of the quantities $z_i$, $L_i$ and $s_i(r)$; we will refer to this assumption as the \emph{local approximation}. The local approximation
can be achieved in many ways, which we will discuss later. Explicitly, the local approximation is $E_{L,i}=\mathcal{E}[z_i,L_i,s_i(r)]$, where 
$\mathcal{E}[z,L,s(r)]$ is the contribution to $E_L$ from any site in the system under consideration which has atomic number $z$ and $L$ electrons with 
a radial distribution described by the function $s(r)$. It is expedient to work with the site-dependent functional 
\begin{equation}
\mathcal{F}_i[Q_i,s_i(r)]\equiv\mathcal{E}[z_i,z_i-Q_i,s_i(r)]
\end{equation}
instead of $\mathcal{E}$, which, like $\mathcal{E}$, gives $E_{L,i}$. Note that $\mathcal{F}_i$ takes $Q_i$ as an argument instead of $L_i$. Furthermore,
since all sites belonging to the same species have the same atomic number, the functionals $\mathcal{F}_i$ are the same for all such sites. 
Applying the local approximation, the total energy $E$ for the system under consideration becomes
\begin{equation}
E=\sum_i\mathcal{F}_i[Q_i,s_i(r)]+E_M,
\end{equation}
where $E_M$ is given by Eqn. \eqref{EM_gen_def}. Note that $E_M$ depends only on the electron density through the values of $Q_i$, and not through the 
functions $s_i(r)$. This is a consequence of the spherical approximation. Now, the ground state $Q_i$ and $s_i(r)$ for all $i$ are those which 
minimize $E$ subject to the following constraints: global charge neutrality (Eqn. \eqref{sumQi=0}), and the validity of Eqn. \eqref{varsigma_const} for 
all $i$. Since $E_M$ is independent of the functions $s_i(r)$, the minimum in $E$ subject to the aforementioned constraints is 
equivalent to the minimum in 
\begin{equation}\label{E=F+EM}
E=\sum_iF_i(Q_i)+E_M
\end{equation}
subject to the single constraint of charge neutrality, where $F_i(Q)$ denotes the minimum value of $\mathcal{F}_i[Q,s(r)]$ over all 
$s(r)$ which obey the analogous equation to Eqn. \eqref{varsigma_const}. The physical significance of $F_i(Q)$ is as follows: $F_i(Q)$
is $E_{L,i}$ if site $i$ contains a net charge $Q$ and the radial distribution of the electron density within the site is allowed to `relax' so to 
obtain its minimum energy configuration. Note that, for all $i$, we no longer need to explicitly impose the constraint of Eqn. \eqref{varsigma_const}
because it is a built-in feature of the function $F_i(Q)$. Note also that, since the functionals $\mathcal{F}_i[Q,s(r)]$ are the same for all 
sites belonging to the same species, then so also are the functions $F_i(Q)$. We will denote the function pertaining to species $X$ as $F_X(Q)$.
Now, each function $F_i(Q_i)$ can be expanded as Taylor series about some charge $\beta_i$. We will choose the quantities $\beta_i$ to be the same for
all sites belonging to the same species: $\beta_i=\beta_X$ for all $i$ belonging to species $X$. \emph{Assuming that the quantities $(Q_i-\beta_i)$ are 
small}, then the Taylor expansions can be shown to yield
\begin{equation}\label{E_gen_def_2}
E=E_0+\frac{1}{2}\sum_ia_i(Q_i-b_i)^2+E_M,
\end{equation}
where
\begin{equation}
a_i= F_i''(\beta_i),
\quad
b_i= \beta_i-\frac{F_i'(\beta_i)}{F_i''(\beta_i)},
\end{equation}
\begin{equation}
E_0\equiv \sum_i\Biggl[F_i(\beta_i)-\frac{1}{2}\frac{F_i'(\beta_i)^2}{F_i''(\beta_i)}\Biggr],
\end{equation}
and $F_i'$ denotes the derivative of $F_i$ with respect to $Q$. As can be seen from Eqns. \eqref{E_gen_def} and \eqref{EL_gen_def}, the
above expression for $E$ is identical to that of the CEFM, except that there is an additional constant term $E_0$ which has no bearing on the values of
$Q_i$ at the minimum in $E$. Note that since $F_i(Q)$ and $\beta_i$ are the same for all sites belonging to the same species, then so also are the
quantities $a_i$ and $b_i$: $a_i$ and $b_i$ take the values $a_X$ and $b_X$ respectively for all $X$ sites as is required within the CEFM.

It is useful to list the assumptions which we have made in the preceding derivation of the CEFM energy function. Our first two assumptions were:
\begin{itemize}
\item That the nuclear positions form an undistorted crystal lattice
\item The spherical approximation.
\end{itemize}
Both of these approximations are commonly employed in \emph{ab initio} electronic structure calculations of disordered alloys. Our remaining
assumptions were:
\begin{itemize}
\item The local approximation
\item That $(Q_i-\beta_i)$ is small for all $i$.
\end{itemize}
Recall that the \emph{Q-V} relations are implicit in the CEFM; they result automatically from minimizing $E$. Therefore any model in which the above 
approximations apply will exhibit the \emph{Q-V} relations. Note that for all $X$ one could \emph{choose} $\beta_X$ to be the mean charge of each species
at the ground state. This is the `best' choice of $\beta_i$ with regards to the validity of the last of the above approximations. In this case the last of 
the above approximations can be restated as follows: that the variance of $Q_i$ for each species is small.

We mentioned earlier that the \emph{local approximation} can be achieved in many ways. We will now elaborate on this point. 
Firstly, it is the case if, for the purposes of evaluating $E_{L,i}$, the region outwith site $i$ is approximated as an effective
medium whose properties somehow reflect the system as a whole. In other words, with regards to calculating $E_{L,i}$ for each site, each site `sees'
the effective medium as its surroundings. It is from this perspective that the CEFM energy function was derived in Ref. \onlinecite{Bruno_2008} within 
the framework of multiple scattering theory. In this case $F_X(Q)$ becomes the local energy of an $X$ site embedded in the effective medium whose charge is
constrained to be $Q$. In fact, since any sensible effective medium will be charge neutral, $E_M=0$ for any system consisting of a site embedded in the 
effective medium, and therefore $F_X(Q)$ is additionally the \emph{total} energy associated with an $X$ site with charge $Q$ embedded in the effective 
medium. 
Note that, if the effective medium is the same for all systems with the same underlying lattice and composition - where we use the term composition to
refer to a particular specification of the concentrations $c_X$ for all $X$ - then so also is $F_X(Q)$ for any particular species $X$, and hence also 
$E_0$, $a_X$ and $b_X$, i.e. the quantities $E_0$, $a_X$ and $b_X$ are \emph{transferable} between systems with the same underlying lattice and
composition. This is true for the SSCPA effective medium, whose construction pays no attention to the specific arrangement of different species on the
underlying lattice. Numerical results also suggest that this is true of the SSLSGF effective medium \cite{Bruno_2008}, but an analytical proof of
this fact has yet to be provided.

Another manner in which the local approximation can be achieved is through the combined use of the Thomas-Fermi and local density approximations,
in which case the contributions to $E_{L,i}$ from the electronic kinetic and exchange-correlation energies depend only on the electron density within
site $i$. The remaining contribution to $E_{L,i}$ is due to the intra-site Coulomb energy of site $i$, which by definition depends only on the contents of 
site $i$. This fact largely explains why in Ref. \onlinecite{Pinski_1998} Pinski was able to reproduce the qualitative aspects of the \emph{Q-V}
relations by using a DFT-based model utilizing the Thomas-Fermi approximation. In Pinski's model all of the approximations listed
earlier are implicit, except for the assumption that $Q_i-\beta_i$ is small for all $i$,
\footnote{In Ref. \onlinecite{Pinski_1998}, Pinski briefly utilized a version of his model in which the spherical approximation does not hold. 
However, we do not consider this here.}
with the local approximation being achieved through the combined use of the Thomas-Fermi and local density approximations. By appealing to the point
made earlier that $\beta_X$ can be chosen to be the mean value of $Q_i$ over all $X$ sites, it follows that it \emph{must} be the case that the \emph{Q-V} 
relations occur in Pinski's model if the variance of $Q_i$ for each species is sufficiently small. 

An alternative approach was described in Ref. \onlinecite{Underwood_2009}. Here, $F_X(Q)$ is chosen to be the energy of a free $X$ ion with charge $Q$. In
this case a logical choice for $\beta_X$ would be the charge of an $X$ atomic core. The quantities $a_X$ and $b_X$ would then apply universally, i.e. they
would transferable between all systems. Furthermore, they could be derived from Hartree-Fock calculations of free ions or experimental values of ionization 
potentials and electron affinities \cite{Underwood_2009}. However, it is optimistic to expect that this approach would result in quantitatively accurate
results.

\section{The CEFM in the non-random approximation}\label{sec:NRA-CEFM}
We will henceforth consider only the case where $a_i=a$ for all $i$. Here, the strength of the local interactions within all sites are the same. 
Following the terminology of Ref. \onlinecite{Drchal_2006}, we will refer to this assumption as the \emph{non-random approximation} (NRA). Furthermore, 
where necessary we will denote the CEFM utilizing the non-random approximation by the abbreviation NRA-CEFM.

\subsection{Accuracy of the non-random approximation}

\begin{figure}
\centering
\includegraphics[height=0.5\textwidth,angle=270]{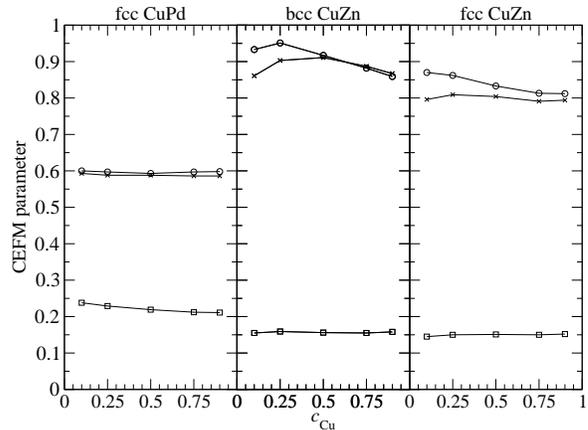}
\caption{CEFM parameters obtained from the LSMS calculations Refs. \onlinecite{Faulkner_1995,Faulkner_1997}. The left, middle and right panels correspond to 
fcc CuPd, bcc CuZn and fcc CuZn random alloys respectively. Values of $a_{\text{Cu}}$ are represented by circles; values of $a_{\text{Pd/Zn}}$ are represented 
by crosses; and values of $(b_{\text{Cu}}-b_{\text{Pd/Zn}})$ are represented by squares. Hartree atomic units are used: $a_{\text{Cu}}$ and $a_{\text{Pd/Zn}}$ are
given in Ha/$e$; and $(b_{\text{Cu}}-b_{\text{Pd/Zn}})$ are given in $e$.}
\label{fig:CEFM_parameters}
\end{figure}

Before investigating the NRA-CEFM in detail, it is worth briefly discussing the loss of accuracy which results from making the NRA. The CEFM parameters obtained
from the LSMS calculations of Refs. \onlinecite{Faulkner_1995,Faulkner_1997} are shown in  Fig. \ref{fig:CEFM_parameters}.
As can be seen from the figure, the differences between $a_{\text{Cu}}$ and $a_{\,\text{Zn/Pd}}$ range from $<1$\% to 10\%. For systems at the lower end of this 
range, it is reasonable to expect that the NRA will yield quantitatively accurate predictions, where by `quantitatively accurate' in this paper we 
mean `in quantitative agreement with DFT calculations utilizing the spherical approximation and a perfect underlying lattice of nuclei'.
While this should not be expected for systems at the upper end of this range, we still expect the NRA to be a useful tool for predicting qualitative trends in 
these systems. We should point out that, while one could use the `general' CEFM to gain at least as accurate results for any particular system as the NRA-CEFM, 
the latter has the advantage that it is significantly simpler, as we will see throughout this paper.

\subsection{Fundamental properties}\label{sec:fundamental_properties}
We begin our investigation of the NRA-CEFM by deriving an explicit expression for $Q_i$.
Recall that Eqns. \eqref{QV_rel} and \eqref{ki_def} hold in the CEFM. Solving these for $Q_i$ yields \cite{Bruno_2007}
\begin{equation}\label{Q=Gab} 
Q_i=\sum_jG_{ij}(a_jb_j+\mu),
\end{equation}
where
\begin{equation}\label{G_H_def}
G\equiv H^{-1},\quad H_{ij}\equiv a_i\delta_{ij}+M_{ij}.
\end{equation}
Setting $a_i=a$ in Eqn. \eqref{Q=Gab}, separating the $j=i$ term from the summation, and using the fact that
\begin{equation}\label{G0}
\zeta\equiv\sum_jG_{ij}=0
\end{equation}
for infinite systems \cite{Drchal_2006,Bruno_2008}, Eqn. \eqref{Q=Gab} becomes
\begin{equation}\label{Q=aGbb}
Q_i=a\sum_{j\neq i}G_{ij}(b_j-b_i).
\end{equation}
Now, as can be seen from Eqn. \eqref{M_def}, $M_{ij}$ takes the same value for all $i$ and $j$ separated by the same distance.
Denoting the $\beta$th nearest neighbor shell of site $i$ as $\beta_i$, and defining the `0th' nearest neighbor shell of site $i$ as the 
set of sites consisting only of site $i$ itself, $M$ can therefore be expressed as follows:
\begin{equation}
M_{ij}=M_{\beta}\quad\text{if $j\in\beta_i$},
\end{equation}
where $M_0=0$, $M_{\beta}=1/R_{\beta}$ for $\beta\geq 1$,%
\footnote{Throughout this paper it should be assumed that generic shell numbers are all $\geq 1$ unless otherwise stated. In other words, one should 
never assume that we are referring to a `0th' nearest neighbor shell. The same applies to Ref. \onlinecite{supp_material}.}
and recall that $R_{\beta}$ is the distance from a site to its $\beta$th nearest neighbor shell.
Now, the matrix $G$ has the same symmetry as the Madelung matrix $M$ within the NRA \cite{Bruno_2007}. Hence for some set of values 
$\lbrace G_0,G_1,G_2,\dotsc\rbrace$,%
\footnote{This can be proven is as follows. Using Eqn. \eqref{G_H_def}, $G$ can be expressed as a power series \cite{lutkepohl}:
$G=a^{-1}(I-a^{-1}M)^{-1}=a^{-1}\sum_{n=0}^{\infty}(-a^{-1}M)^n$, where $I$ denotes the identity matrix. It can then be shown by induction that 
all terms in the series have the same symmetry as $M$, and hence so also must $G$.}
\begin{equation}
G_{ij}=G_{\beta}\quad\text{if $j\in\beta_i$}.
\end{equation}
With this in mind, separating the summation in Eqn. \eqref{Q=aGbb} into contributions from each shell of $i$ gives
\begin{equation}
Q_i=a\sum_{\beta=1}^{\infty}G_{\beta}\sum_{j\in\beta_i}(b_j-b_i).
\end{equation}
Splitting the summation over $j\in\beta_i$ into contributions from each species $Y$ present in the alloy gives
\begin{equation}\label{QX_transfer}
Q_i=\Lambda\sum_Yb_{YX}\sum_{\beta=1}^{\infty}g_{\beta}N_{iY\beta}\quad \text{if $i\in X$},
\end{equation}
where $N_{iY\beta}$ is the number of sites belonging to species $Y$ which are in shell $\beta$ of $i$, we have introduced the quantities
\begin{equation}\label{Lambda_gbeta_def}
\Lambda\equiv aG_0, \quad g_{\beta}\equiv G_{\beta}/G_0
\end{equation}
and
\begin{equation}\label{bYX_def}
b_{YX}\equiv b_Y-b_X,
\end{equation}
and without loss of generality we have chosen site $i$ to belong to species $X$. From Eqn. \eqref{G0} it follows that the values of $g_{\beta}$ are 
constrained to obey the relation
\begin{equation}\label{gbetaZbeta=-1}
\sum_{\beta=1}^{\infty}g_{\beta}Z_{\beta}=-1,
\end{equation}
where $Z_{\beta}$ denotes the number of sites in shell $\beta$ of any site in the system. This will be used later.

Eqn. \eqref{QX_transfer} allows us to interpret the charge distribution throughout the system under consideration in terms of charge transfer 
between all pairs of sites: from each $Y$ site in shell $\beta$ of an $X$ site $i$, the $X$ site receives a quantity of charge
$\Lambda b_{YX}g_{\beta}$. Conversely, the $X$ site itself donates a quantity of charge $\Lambda b_{XY}g_{\beta}$ to each $Y$ site in shell $\beta$.
Since $b_{XY}=-b_{YX}$, the charge donated to the $X$ site from the $Y$ site, and the charge donated to the $Y$ site from the $X$ site, 
are equal in magnitude but opposite in sign. Therefore a quantity of charge $|\Lambda b_{XY}g_{\beta}|$ can be regarded as being transferred 
directly between each pair of $X$ and $Y$ sites separated by $R_{\beta}$. Interestingly, since $b_{XX}=0$, charge is transferred only between
pairs of \emph{unlike} sites. More generally, the species-dependence of the amount of charge transferred between an $X$ and $Y$ site enters
entirely through the difference in their bare charges $b_{XY}$: the higher the difference; the higher the amount of charge transferred. We will 
re-examine this point in the next section after calculating $\Lambda$ and $g_{\beta}$ for a wide range of systems.

The fact that the values of $Q_i$ in the NRA-CEFM can be understood in terms of charge transfer between pairs of unlike sites is reminiscent of the 
MLCM. For binary alloys, Eqn. \eqref{QX_transfer} can be written in the form of Eqn. \eqref{Q_MLCM} with $\beta_{\text{max}}=\infty$ and
\begin{equation}\label{lambdabeta_CEF}
\lambda_{\beta}=\frac{1}{2}\Lambda b_{AB}g_{\beta}.
\end{equation}
Hence the NRA-CEFM for binary alloys is a particular case of the MLCM. In fact, since the \emph{Q-V} relations hold exactly in the NRA-CEFM, 
the model is equivalent to the OLCM (with $\beta_{\text{max}}=\infty$) for binary alloys - where recall that the OLCM is the
particular case of the MLCM in which the free parameters are constrained to give \emph{Q-V} relations as closely as is possible. The assumptions
described in Sec. \ref{sec:underlying_approx}, in addition to the non-random approximation, therefore also underpin the OLCM. We demonstrated earlier
that these assumptions should yield at least qualitatively accurate results; the same therefore applies to the OLCM. Thus we have put
the OLCM on a firm theoretical footing, whereas before it was somewhat of an \emph{ad hoc} rule which seemed to reproduce order-$N$ results
given the correct parameters. Furthermore, in Eqn. \eqref{QX_transfer} we have discovered the generalization of the OLCM charge law which can be
applied to alloys containing any number of species - not just two.

The values of $V_i$ can be determined using Eqns. \eqref{QX_transfer}, \eqref{QV_rel} and \eqref{ki_def} - once $\mu$ has been determined. We will 
now derive an expression for $\mu$. Substituting Eqn. \eqref{Q=Gab} into Eqn. \eqref{sumQi=0} gives
\begin{equation}
\mu=-\frac{\sum_{i,j}G_{ij}a_jb_j}{\sum_{ij}G_{ij}}.
\end{equation}
This becomes
\begin{equation}
\mu=-a\frac{\sum_jb_j\zeta}{\sum_i\zeta}
\end{equation}
after setting $a_i=a$. Now, taking the limit $\zeta\to 0$ amounts to applying Eqn. \eqref{G0}; doing so gives
\begin{equation}\label{mu}
\mu=-a\langle b\rangle,
\end{equation}
where $\langle b\rangle$ denotes the mean value of $b_i$ over all $i$.

\section{Screening}\label{sec:screening}
Henceforth we will assume that $a$ and the values of $b_X$ are known for the system under consideration. These could have been extracted
from the results of \emph{ab initio} calculations, or determined by some other means. With this information, however, 
we still do not know the values of $g_{\beta}$ and $\Lambda$. These are the `geometric factors' which were mentioned at the end of
Sec. \ref{sec:intro}. Knowledge of these quantities is required before Eqn. \eqref{QX_transfer}, as well as those derived later, can be used 
in practice. In this section, we will calculate them for a wide range of systems. In doing this, we will learn much about the nature of the 
screening in the NRA-CEFM.

Consider the matrix $aG$. Using Eqns. \eqref{M_def} and \eqref{G_H_def}, it can be expressed as
\begin{equation}
aG=\biggl(I+\frac{1}{aR_{\text{WS}}}M^{R_{\text{WS}}=1}\biggr)^{-1},
\end{equation}
where $M^{R_{\text{WS}}=1}$ is the Madelung matrix for the same lattice as the system under consideration, but with unit Wigner-Seitz radius 
$R_{\text{WS}}$. Note that $aG$ for any particular system depends only on: its lattice \emph{type}%
\footnote{By a lattice's \emph{type} we mean, for example, fcc, bcc, sc. Note that two lattices with the same type can have different values of
$R_{\text{WS}}$.}
(through $M^{R_{\text{WS}}=1}$), and the value of $aR_{\text{WS}}$. Now, as can be seen from Eqn. \eqref{Lambda_gbeta_def}, $\Lambda$ and $g_{\beta}$
can be derived from the elements of $aG$ as follows: the diagonal elements of $aG$ are all $\Lambda$; $g_{\beta}=aG_{ij}/\Lambda$ for any pair
of sites $i$ and $j\in\beta_i$. Therefore $\Lambda$ and $g_{\beta}$ depend only on $aR_{\text{WS}}$ and the lattice type, and
hence we need only calculate them once for each combination of $aR_{\text{WS}}$ and lattice type. \emph{No analogous simplification can be made
in the general CEFM.} This stems from the fact that, in the general CEFM, $H$ depends on the values of $a_i$, and hence $G$ is different for 
different systems even if they have the same lattice type, value of $R_{\text{WS}}$, and set of values $a_X$. In this sense the NRA-CEFM is far
more practicable than the general CEFM.

In a moment we present asymptotic expressions and numerical results which describe how $\Lambda$ and $g_{\beta}$ vary with $aR_{\text{WS}}$ for
various lattice types. However, it is instructive to first understand the physical significance of these quantities. We begin with
$aR_{\text{WS}}$. Recall that $a$ determines the strength of the local interactions which act to keep the site charges $Q_i$ at their bare
values $b_i$. The analogous quantity for the inter-site Coulomb interactions is $1/R_{\text{WS}}$: smaller inter-site separations mean stronger 
inter-site Coulomb interactions. With this in mind, it can be seen that $aR_{\text{WS}}$ is a dimensionless quantity which determines the strength
of the local interactions \emph{relative} to the strength of the inter-site Coulomb interactions: the higher the value of $aR_{\text{WS}}$, the 
more important the local interactions are, and the less important the inter-site Coulomb interactions are, in determining the values of $Q_i$ 
which minimize $E$. The physical significance of $\Lambda$ and $g_{\beta}$ can be related to the nature of the screening. In Ref. 
\onlinecite{Drchal_2006} it was shown that the strength of the local interactions at site $i$, renormalized by the electrostatic interactions 
with the rest of the system, is $a^{\text{scr}}_i=1/G_{ii}$ in the general CEFM. From Eqn. \eqref{Lambda_gbeta_def} it therefore follows that
$\Lambda=a/a^{\text{scr}}$, i.e. $\Lambda$ is a measure of the \emph{amount of screening} which occurs in the system. In the absence of screening 
$\Lambda=1$; in the presence of screening $\Lambda>1$. In Ref. \onlinecite{Drchal_2006} it was also shown that if $Q_i$ is perturbed by a certain
amount $\delta Q_i$, and the charges on all other sites in the system are allowed to `relax' so to minimize the total energy of the system, then
the resulting change in $Q_j$ is given by
\begin{equation}
\delta Q_j=\frac{G_{ji}}{G_{ii}}\delta Q_i
\end{equation}
in the general CEFM. From Eqn. \eqref{Lambda_gbeta_def} it can be seen that
\begin{equation}
\delta Q_j=g_{\beta}\delta Q_i
\end{equation}
for $j\in\beta_i$ in the NRA-CEFM. The above equation reveals that the values of $g_{\beta}$ describe the radial distribution of screening charge
around a charge perturbation: for a perturbation $\delta Q_0$, the induced charge on sites at distance $R_{\beta}$ from the perturbation is
$\delta Q_{\beta}=g_{\beta}\delta Q_0$. Note that
\begin{equation}
g_{\beta}=\delta Q_{\beta}/\delta Q_0,
\end{equation}
and hence the values of $g_{\beta}$ can be directly compared to the $\delta Q_{\beta}/\delta Q_0$ obtained by RS (Eqn. \eqref{us_function}). This 
will be done later.

\subsection{Asymptotic expressions}
The following asymptotic expressions for $\Lambda$ and $g_{\beta}$, apply for the limit $aR_{\text{WS}}\to\infty$:\cite{supp_material}
\begin{equation}\label{aG0_analytic}
\Lambda= \frac{aR_{\text{WS}}(1+\nu)}{aR_{\text{WS}}(1+\nu)-\nu}
\end{equation}
and
\begin{equation}\label{gbeta_analytic}
g_{\beta}=-\Biggl[\frac{\nu^2e^{\nu}}{3(1+\nu)}\Biggr]
\frac{\exp(-\nu R_{\beta}/R_{\text{WS}})}{R_{\beta}/R_{\text{WS}}},
\end{equation}
where
\begin{equation}\label{nu_def}
\nu=\sqrt{\frac{3}{aR_{\text{WS}}-1.5}}.
\end{equation}
In the limit $aR_{\text{WS}}\to\infty$ the values of $Q_i$ will be determined solely by the local interactions. This can be realized by setting $E_M=0$ 
in Eqn. \eqref{E_gen_def}, in which case minimization of $E$ subject to Eqn. \eqref{sumQi=0} yields 
\begin{equation}
Q_i=b_i-\langle b\rangle
\end{equation} 
for all $i$, where we have used Eqn. \eqref{mu}. In other words, all sites assume their bare charges, with an additional amount of charge 
$-\langle b\rangle$ added to all sites to enforce global charge neutrality. 
We can learn several things from Eqns. \eqref{aG0_analytic} and \eqref{gbeta_analytic}. Consider Eqn. \eqref{aG0_analytic} first. As can be seen
from Eqn. \eqref{nu_def}, $\nu\to 0$ as $aR_{\text{WS}}\to\infty$; therefore $\Lambda\to 1$ as $aR_{\text{WS}}\to\infty$. Recall that $\Lambda$ is
a measure of the amount of screening. Therefore the amount of screening vanishes as $aR_{\text{WS}}\to\infty$. Consider now Eqn. 
\eqref{gbeta_analytic}. This can be written in the form
\begin{equation}
g_{\beta}=u(R_{\beta}/R_{\text{WS}}),
\end{equation}
where $u$ is a function independent of lattice type, i.e. it is a \emph{universal function}. With this in mind, we see that the above
equation describes the qualitative aspect of universal screening observed by RS. To restate: universal screening is implicit in the NRA-CEFM
in the limit $aR_{\text{WS}}\to\infty$. We will elaborate on this result in the next subsection. Another point worth mentioning regarding Eqn. 
\eqref{gbeta_analytic} is that the function $u$ is of Yukawa form. Interestingly, the analogous parameters to $g_{\beta}$ in the MLCM and OLCM, 
when parameterized using LSMS results in Refs. \onlinecite{Wolverton_1996} and \onlinecite{Underwood_2009} respectively, were also 
observed to vary with $R_{\beta}/R_{\text{WS}}$ in this manner. However, the dependence of $g_{\beta}$ with $R_{\beta}/R_{\text{WS}}$ observed in 
these studies is empirical since the free parameters in the MLCM and OLCM had in both cases been fit to LSMS data. By contrast, Eqn. 
\eqref{gbeta_analytic} is an analytical result valid in the limit of large $aR_{\text{WS}}$.

\subsection{Numerical results}\label{sec:numerical}

\begin{figure}
\centering
\includegraphics[height=0.5\textwidth,angle=270]{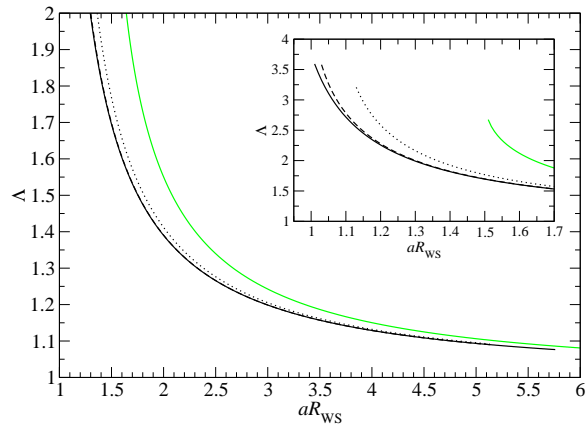}
\caption{(Color online) $\Lambda$ vs. $aR_{\text{WS}}$. The black solid, dashed and dotted curves correspond to the results of the numerical
calculations for the fcc, bcc and sc lattices respectively. The green (gray) solid curve corresponds to the predictions of Eqn. 
\eqref{aG0_analytic}.}
\label{fig:aG0_vs_aRWS}
\end{figure}

\begin{figure}
\centering
\includegraphics[width=0.5\textwidth]{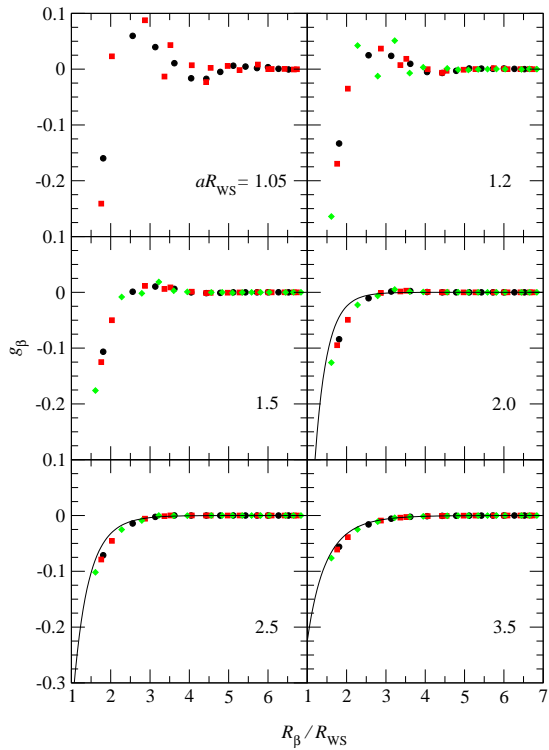}
\caption{(Color online) $g_{\beta}$ vs. $R_{\beta}/R_{\text{WS}}$ for various values of $aR_{\text{WS}}$. Each panel corresponds to a particular 
$aR_{\text{WS}}$, whose value is indicated. The black circles, red (dark gray) squares and green (light gray) diamonds correspond to the results 
of the numerical calculations for the fcc, bcc and sc lattices respectively. The curves correspond to the predictions of Eqn. \eqref{gbeta_analytic}.}
\label{fig:gbeta_vs_Rbeta}
\end{figure}

We have numerically determined $\Lambda$ and $g_{\beta}$ as a function of $aR_{\text{WS}}$ for the fcc, bcc and sc lattices. Details of
the procedure, as well as tables of the results, can be found in Ref. \onlinecite{supp_material}.
Fig. \ref{fig:aG0_vs_aRWS} shows how $\Lambda$ varies with $aR_{\text{WS}}$, and Fig. \ref{fig:gbeta_vs_Rbeta} shows the values
of $g_{\beta}$ plotted against $R_{\beta}/R_{\text{WS}}$ at selected values of $aR_{\text{WS}}$. Note that in both figures the numerical
results tend to the asymptotic predictions of Eqns. \eqref{aG0_analytic} and \eqref{gbeta_analytic} as $aR_{\text{WS}}$ is increased.
We will begin by discussing Fig. \ref{fig:aG0_vs_aRWS}. This figure reveals that, for all lattice types considered, $\Lambda$ decreases 
monotonically to 1 as $aR_{\text{WS}}$ increases. Recall that $\Lambda$ is a measure of the amount of screening. Therefore the amount of screening
decreases monotonically with $aR_{\text{WS}}$. This makes sense. Earlier, we mentioned that the quantity $aR_{\text{WS}}$ determines the strength of the 
local interactions relative to the strength of the inter-site (Coulomb) interactions. Bearing in mind that screening arises wholly as an attempt to 
reduce the energy associated with inter-site interactions, it follows that for high values of $aR_{\text{WS}}$, i.e. weak inter-site interactions, the 
amount of screening will be low. Conversely, for low values of $aR_{\text{WS}}$, the amount of screening will be high.

Consider now Fig. \ref{fig:gbeta_vs_Rbeta}. Recall that $g_{\beta}$ describes the radial distribution of charge which screens a charge
perturbation. The tendency for the screening to decrease as $aR_{\text{WS}}$ increases is reflected in this figure: at high values of 
$aR_{\text{WS}}$ the radial distribution of screening charge is `flat', which corresponds to a less localised screening charge distribution; while for
lower values of $aR_{\text{WS}}$ the radial distribution is increasingly `skewed' towards the central site, which corresponds to more localised
screening charge distribution. Interestingly, while the values of $g_{\beta}$ increase to 0 monotonically for large values of $aR_{\text{WS}}$, they exhibit 
oscillations for lower values of $aR_{\text{WS}}$. Furthermore, the oscillations become increasingly violent as $aR_{\text{WS}}$ 
is decreased. The same phenomenon was observed by Drchal \emph{et al.} \cite{Drchal_2006} with regards to the screened Coulomb interactions.
This can be understood in terms of the Gibb's-phenomenon. Eqn. \eqref{gbeta_analytic} describes how $g_{\beta}$ depends on 
$R_{\beta}/R_{\text{WS}}$ for large values of $aR_{\text{WS}}$, and is of Yukawa form. The range of the $g_{\beta}$ vs.
$R_{\beta}/R_{\text{WS}}$ curve of Eqn. \eqref{gbeta_analytic} is determined by the value of $aR_{\text{WS}}$: the higher
$aR_{\text{WS}}$ is, the longer its range in real space, and hence the shorter its range in reciprocal space. Now, due to the 
discreteness of the lattice, there is an upper limit to the wavevectors which can be represented upon it. 
This fact is only important for low values of $aR_{\text{WS}}$. Here, the Yukawa function has strong Fourier components above the lattice's upper limit,
the result of which is that these components are `cut out' of the Yukawa function, leaving oscillations in the $g_{\beta}$ vs.
$R_{\beta}/R_{\text{WS}}$ curve in real space. Note that the upper limit is different for different lattice types at the same $R_{\text{WS}}$, 
which is why the $g_{\beta}$ vs. $R_{\beta}/R_{\text{WS}}$ curves at low values of $aR_{\text{WS}}$ differ for different lattice types.
Hence the universality in $g_{\beta}$ vs. $R_{\beta}/R_{\text{WS}}$ at high $aR_{\text{WS}}$, described by Eqn. \eqref{gbeta_analytic},
increasingly breaks down as $aR_{\text{WS}}$ is decreased. This is borne out in Fig. \ref{fig:gbeta_vs_Rbeta}. 
Consider the panel in the figure corresponding to $aR_{\text{WS}}=1.5$. Here we see that, while the $(R_{\beta}/R_{\text{WS}},g_{\beta})$
points for the fcc and bcc lattices still appear to lie upon the same curve, the points for the sc lattice do not. Decreasing
$aR_{\text{WS}}$ below this only increases the deviation of the sc points from the fcc/bcc curve. At $aR_{\text{WS}}=1.2$ the deviation
is significant. Further decreases in $aR_{\text{WS}}$ similarly cause the bcc curve to `break away' from the fcc curve, as can be
seen from the $aR_{\text{WS}}=1.05$ panel in the figure.

In light of our calculations of $\Lambda$ and the values of $g_{\beta}$, let us now briefly reconsider charge transfer. As was mentioned in Sec. 
\ref{sec:fundamental_properties}, the values of $Q_i$ in the NRA-CEFM can be understood as resulting from charge transfer between all pairs of sites as 
follows: each $X$ site transfers an amount $\Lambda b_{XY}g_{\beta}$ to each $Y$ site at $R_{\beta}$ from the $X$ site. As can be seen from Fig. 
\ref{fig:gbeta_vs_Rbeta}, for a given lattice type there is a threshold value of $aR_{\text{WS}}$ above which, for all practical purposes, $g_{\beta}<0$ 
for all $\beta$. Noting that $\Lambda$ is always positive - as can be seen from Fig. \ref{fig:aG0_vs_aRWS} - it follows that above this threshold both
the species-dependence and the \emph{direction} of the charge transferred between an $X$ site and a $Y$ site is completely determined by $b_{XY}$: if 
$b_X>b_Y$ then $X$ sites gain charge from $Y$ sites; if $b_X<b_Y$ then $X$ sites lose charge to $Y$ sites. Furthermore, the larger $|b_{XY}|$, the larger 
the amount of charge transferred between $X$ and $Y$ sites at a given distance. For such values of $aR_{\text{WS}}$ one can therefore regard charge transfer
as being governed by an electronegativity-like relationship, with $b_X$ playing the role of the electropositivity of species $X$. However, this viewpoint
becomes less useful for low values of $aR_{\text{WS}}$, where the direction of charge transfer between an $X$ and a $Y$ site varies with their separation, 
as can be seen from the fact that more and more values of $g_{\beta}$ become positive as $aR_{\text{WS}}$ is decreased.
For the fcc CuPd, bcc CuZn and fcc CuZn random alloy systems considered in Ref. \onlinecite{Faulkner_1997} the LSMS values of $aR_{\text{WS}}$ are 1.6, 2.4 
and 2.2 respectively, where in this paper $a$ derived from LSMS results is always taken to be the concentration-weighted average of $a_A$ and $a_B$, i.e. 
$a=c_Aa_A+c_Ba_B$. For the latter two of these systems $aR_{\text{WS}}$ lies in the range where charge transfer can be considered to be governed by an
electronegativity-like relationship.

\begin{figure}
\centering
\includegraphics[height=0.5\textwidth]{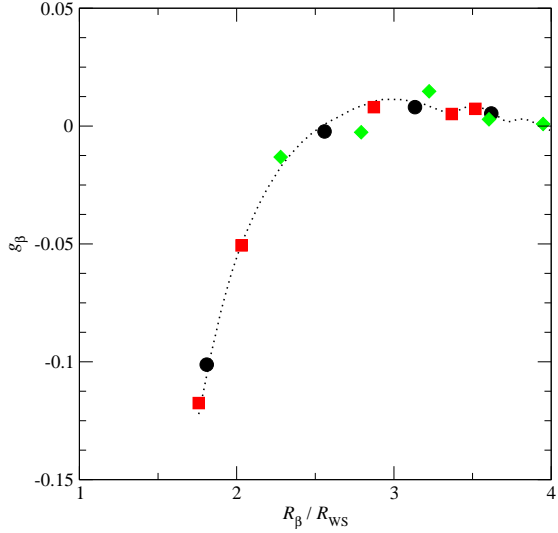}
\caption{(Color online) The values of $g_{\beta}$ plotted against $R_{\beta}/R_{\text{WS}}$ for $aR_{\text{WS}}=1.6$. The black circles, red (dark gray)
squares and green (light gray) diamonds correspond to the fcc, bcc and sc lattices respectively. The dotted curve corresponds to the results of RS 
(Ref. \onlinecite{Ruban_2002}).}
\label{fig:gbeta_us}
\end{figure}

To conclude this section we will compare our results to those of RS. Recall that RS, using the SSLSGF method, observed that for all systems 
$\delta Q_{\beta}/\delta Q_0$ vs. $R_{\beta}/R_{\text{WS}}$ is a universal curve. They also observed that Eqns. \eqref{us_aR} and \eqref{us_kR} applied for 
all species in all systems. We expect our results to be in quantitative agreement with those of RS, because the approximations which underpin the
NRA-CEFM are either implicit in the SSLSGF method or can be justified \emph{a posteriori} from RS's results themselves. To elaborate, the first three 
approximations listed in Sec. \ref{sec:underlying_approx} are implicit in the SSLSGF method, where the local approximation is achieved through the use
of an effective medium - as described in Sec. \ref{sec:intro}. Now, it is not obvious whether or not the final approximation applies, namely, that for some 
choice of $\beta_i$, $(Q_i-\beta_i)$ is small for all $i$. However, it was pointed out in Sec. \ref{sec:underlying_approx} that if all of the 
listed approximations are satisfied, then the \emph{Q-V} relations \emph{must} hold. The fact that the \emph{Q-V} relations hold to a high degree of
accuracy in RS's results \cite{Ruban_2002} therefore implies \emph{a posteriori} that the values of $(Q_i-\beta_i)$ for all $i$ must be small. There is 
one further approximation which underpins the NRA-CEFM in addition to those listed in Sec. \ref{sec:underlying_approx}, namely, that the values of $a_X$ 
are the same for all species in a given system, i.e. the non-random approximation. This is also borne out in RS's results, as can be seen from Eqn. 
\eqref{us_aR}. Note that the exact reasons \emph{why} the values of $a_X$ are the same for all species in a given system, and are such that 
$a_XR_{\text{WS}}\approx 1.6$ for all systems, are still unclear and require further investigation.
In Fig. \ref{fig:gbeta_us}, our results for $aR_{\text{WS}}=1.6$ are compared to the curve obtained by RS. From the figure, it can be seen that our fcc
and bcc points agree well with the curve - as expected. However, the points corresponding to the sc lattice do not. This is because RS only considered
systems with the fcc, bcc or bct structure: they did not consider systems with the sc structure. An alloy exhibiting the sc structure would therefore
`break' the universality described by RS. In fact, it should be noted that the universality breaks down anyway if one uses a more accurate model that that
provided by the SSLSGF method. This is clear from the wide range of $aR_{\text{WS}}$ obtained from the LSMS method, as listed in the preceding paragraph.

\section{Analytical results}\label{sec:analytical}
We will now use Eqn. \eqref{QX_transfer} to derive analytical expressions for various physical quantities. 
We begin by introducing the quantities $p^{XY}_{\beta}$ and $p^{XYZ}_{\beta\gamma}$, which will appear in the expressions derived 
below. The physical significance of the quantities $p^{XY}_{\beta}$ is as follows: $p^{XY}_{\beta}$ is the probability that a site $j$ at distance
$R_{\beta}$ from an $X$ site $i$ belongs to species $Y$, where both $i$ and $j$ are selected at random subject to the aforementioned constraints. It
should be noted that the values of $p^{XY}_{\beta}$ for a given $X$ and $\beta$ are constrained to obey the equation
\begin{equation}
\sum_Yp^{XY}_{\beta}=1,
\end{equation}
i.e. the probability that a site at distance $R_{\beta}$ of a randomly selected $X$ site belongs to \emph{any} species is 1. Furthermore, if the system 
under consideration is a binary alloy, then the values of $p^{XY}_{\beta}$ are related to $\alpha_{\beta}$ - the Warren-Cowley short range order parameter
pertaining to sites separated by $R_{\beta}$ - by the following equations \cite{Cowley_1960}:
\begin{equation}
p^{AA}_{\beta}=c_A+c_B\alpha_{\beta},\quad p^{AB}_{\beta}=c_B(1-\alpha_{\beta}),
\end{equation}
\begin{equation}
p^{BA}_{\beta}=c_A(1-\alpha_{\beta}),\quad p^{BB}_{\beta}=c_B+c_A\alpha_{\beta}.
\end{equation}
In a similar vein to $p^{XY}_{\beta}$, $p^{XYZ}_{\beta\gamma}$ is the probability that sites $j$ and $k$, which are distinct and at distances $R_{\beta}$
and $R_{\gamma}$ respectively from an $X$ site $i$, belong to species $Y$ and $Z$ respectively - where all sites are selected randomly subject to the
aforementioned constraints. Note that by saying \emph{distinct} we have disallowed the case where $j$ and $k$ are the same site (which could only
occur if $\gamma=\beta$). Similarly to $p^{XY}_{\beta}$, the values of $p^{XYZ}_{\beta\gamma}$ for a given $X$, $\beta$ and $\gamma$ are constrained to 
obey the equation
\begin{equation}
\sum_{Y,Z}p^{XYZ}_{\beta\gamma}=1.
\end{equation}

In what follows we will use the following results:\cite{supp_material}
\begin{equation}\label{mean_NiYbeta}
\langle N_{iY\beta}\rangle_X=Z_{\beta}p_{\beta}^{XY}
\end{equation}
and
\begin{equation}\label{covar_NN}
\begin{split}
\Cov(N_{iY\beta},N_{iZ\gamma})_X=&
Z_{\beta}Z_{\gamma}(p^{XYZ}_{\beta\gamma}-p^{XY}_{\beta}p^{XZ}_{\gamma})\\
&+\delta_{\beta\gamma}Z_{\beta}(\delta_{YZ}p^{XY}_{\beta}-p^{XYZ}_{\beta\beta}),
\end{split}
\end{equation}
where $\langle P_i\rangle_X$ denotes the mean value of $P_i$ over all $X$ sites $i$, and $\Cov(P^{(1)}_i,P^{(2)}_i)_X$ denotes the covariance of 
$P^{(1)}_i$ and $P^{(2)}_i$ over all $X$ sites $i$.

\subsection{Mean and variance of $Q_i$ and $V_i$}
Taking the mean of Eqn. \eqref{QX_transfer} over $X$ sites yields
\begin{equation}
\langle Q\rangle_{X}=\Lambda\sum_Yb_{YX}\sum_{\beta=1}^{\infty}g_{\beta}\langle N_{iY\beta}\rangle_X
\end{equation}
after exploiting the linearity of the mean. This becomes
\begin{equation}\label{QX_mean}
\langle Q\rangle_{X}=\Lambda\sum_Yb_{YX}\sum_{\beta=1}^{\infty}g_{\beta}Z_{\beta}p^{XY}_{\beta}
\end{equation}
after using Eqn. \eqref{mean_NiYbeta}. 

Taking the variance of Eqn. \eqref{QX_transfer} over $X$ sites yields
\begin{equation}\label{proof_VarQ}
\Var(Q)_X=\Lambda^2\sum_{Y,Z}b_{YX}b_{ZX}\sum_{\beta,\gamma=1}^{\infty}g_{\beta}g_{\gamma}\Cov(N_{iY\beta},N_{iZ\gamma})_X
\end{equation}
after exploiting the bilinearity of the covariance. This becomes
\begin{widetext}
\begin{equation}\label{QX_Var}
\Var(Q)_X=\Lambda^2\Biggl[
\sum_{Y}b_{YX}^2\sum_{\beta=1}^{\infty}g_{\beta}^2Z_{\beta}p^{XY}_{\beta}
-\sum_{Y,Z}b_{YX}b_{ZX}\sum_{\beta=1}^{\infty}g_{\beta}^2Z_{\beta}p^{XYZ}_{\beta\beta}
+\sum_{Y,Z}b_{YX}b_{ZX}\sum_{\beta,\gamma=1}^{\infty}g_{\beta}g_{\gamma} 
Z_{\beta}Z_{\gamma}(p^{XYZ}_{\beta\gamma}-p^{XY}_{\beta}p^{XZ}_{\gamma})
\Biggr]
\end{equation}
\end{widetext}
after using Eqn. \eqref{covar_NN}.

From $\langle Q\rangle_X$ and $\Var(Q)_X$, $\langle V\rangle_X$ and $\Var(V)_X$ can be calculated using the following equations:
\begin{equation}\label{VX_mean}
\langle V\rangle_X=-a\langle Q\rangle_X+a(b_X-\langle b\rangle)
\end{equation}
and
\begin{equation}\label{VX_Var}
\Var(V)_X=a^2\Var(Q)_X.
\end{equation}
The former equation follows from taking the mean of Eqn. \eqref{QV_rel} for $X$ sites and using Eqn. \eqref{mu}; the latter follows from taking
the variance of Eqn. \eqref{QV_rel}.

\subsection{Core level shifts}\label{subsec:CLSs}
In the ESCA potential model the metal-alloy \emph{initial state} core level shift for an $X$ site $i$ is given by
\cite{Williams_1978,Gelius_1974,Weinert_1995}
\begin{equation}\label{Vtot_ESCA}
\Delta E_i^{\text{B}}=\frac{Q_i}{r^{\text{eff}}_i}+V_i+Q^{\text{val,metal}}_X\Delta(1/r^{\text{eff}}_i),
\end{equation}
where $r^{\text{eff}}_i$ is the effective radius from nucleus $i$ at which the valence charge within site $i$ can be considered to reside with regards
to the total electrostatic potential at the nucleus, $Q^{\text{val,metal}}_X$ is the valence charge on a site in a pure $X$ metal, and 
$\Delta(1/r^{\text{eff}}_i)$ is the shift in $1/r^{\text{eff}}_i$ relative to the corresponding value for a site in a pure $X$ metal. 
\emph{Assuming that $r^{\text{eff}}_i$ takes the same value $r^{\text{eff}}_X$ for all $X$ sites in the alloy under consideration}, the last term becomes
a system- and species-dependent constant. We will denote this constant as $\Theta_X$. With this in mind, and using Eqns. \eqref{QV_rel} and 
\eqref{mu}, it follows that
\begin{equation}\label{VtotX}
\Delta E_i^{\text{B}}=\biggl(\frac{1}{r^{\text{eff}}_X}-a\biggr)Q_i + a\bigl(b_X-\langle b\rangle\bigr) + \Theta_X
\end{equation}
for $X$ sites in the NRA-CEFM. Taking the mean and variance of this equation over all $X$ sites gives
\begin{equation}\label{VtotX_mean}
\langle\Delta E^{\text{B}}\rangle_X=\biggl(\frac{1}{r^{\text{eff}}_X}-a\biggr)\langle Q\rangle_X + a\bigl(b_X-\langle b\rangle\bigr) + \Theta_X
\end{equation}
and
\begin{equation}\label{VtotX_Var}
\Var(\Delta E^{\text{B}})_X=\biggl(\frac{1}{r^{\text{eff}}_X}-a\biggr)^2\Var(Q)_X
\end{equation}
respectively. $\Var(\Delta E^{\text{B}})_X$ is a measure of the initial state core level \emph{disorder broadening} for a given $X$ core level.

It should be emphasized that the NRA-CEFM is by no means limited to evaluation of the initial state contribution to core level shifts. It is
possible to use the model to derive expressions for core level shifts which include final state contributions. This can be done by evaluating the
difference in $E$ as a result of the transformation $b_i\to b_i^*$, where $b_i^*$ is the bare charge associated with site $i$ if its atomic core
is ionized, and $i$ is the site whose core level shifts we are interested in. A similar procedure could be used to derive expressions for Auger
kinetic energy shifts. However, this is beyond the scope of this paper.

\subsection{Energies}
In Ref. \onlinecite{Drchal_2006} Drchal \emph{et al.} derive expressions for $E_L$, $E_M$ and $E$ in terms of $\langle Q\rangle_X$, $\langle V\rangle_X$ 
and $\Var(Q)_X$ for binary alloys in the general CEFM. Using a similar procedure, we find that the analogous intensive energies $\tilde{E}_L$,
$\tilde{E}_M$ and $\tilde{E}$ for any alloy within the NRA are given by
\begin{equation}\label{EL}
\tilde{E}_L=\frac{1}{2}a\sum_Xc_X\Bigl[\bigl(b_X-\langle Q\rangle_X\bigr)^2+\Var(Q)_X\Bigr],
\end{equation}
\begin{equation}\label{EM}
\tilde{E}_M=\frac{1}{2}a\sum_Xc_X\Bigl[\bigl(b_X-\langle Q\rangle_X\bigr)\langle Q\rangle_X-\Var(Q)_X\Bigr]
\end{equation}
and
\begin{equation}\label{E}
\tilde{E}=\frac{1}{2}a\sum_Xc_Xb_X\Bigl[b_X-\langle Q\rangle_X\Bigr],
\end{equation}
where recall that $c_X$ denotes the global concentration of $X$ sites.

\subsection{Random alloys}
We will now apply the above equations to random alloys, for which $p^{XY}_{\beta}=c_Y$ and $p^{XYZ}_{\beta\gamma}=c_Yc_Z$. Note that the concentrations obey
\begin{equation}\label{sumcY=1}
\sum_Yc_Y=1,
\end{equation}
something which we will exploit multiple times below. For random alloys, Eqns. \eqref{QX_mean} and \eqref{QX_Var} become
\begin{equation}\label{QXR_mean}
\langle Q\rangle_X =\Lambda(b_X-\langle b\rangle)
\end{equation}
and
\begin{equation}\label{QXR_Var}
\Var(Q)_X=\Lambda^2\omega\Var(b),
\end{equation}
where $\Var(b)$ is the variance in the values of $b_i$ for all $i$, and
\begin{equation}\label{omega_def}
\omega\equiv\sum_{\beta=1}^{\infty}g_{\beta}^2Z_{\beta}.
\end{equation}
The values of $\omega$ calculated using the $g_{\beta}$ obtained from our numerical calculations are tabulated in Ref.
\onlinecite{supp_material}. In deriving Eqn. \eqref{QXR_mean} we have used Eqn. \eqref{gbetaZbeta=-1} and the fact that
\begin{equation}
\sum_Yb_{YX}c_Y=\sum_Yb_Yc_Y-b_X\sum_Yc_Y=\langle b\rangle-b_X,
\end{equation}
which itself follows from the definition of $b_{YX}$ and Eqn. \eqref{sumcY=1}.
In deriving Eqn. \eqref{QXR_Var} we have exploited the fact that any variance is invariant under any rigid transformation of the
random variables under consideration, and hence $\Var(b)$ is invariant under the transformation $b_i\to b_i-b_X$; therefore
\begin{equation}\label{Var_b}
\begin{split}
\Var(b)=&\langle b^2\rangle-\langle b\rangle^2 \\
=&\sum_Yb_Y^2c_Y-\biggl(\sum_Yb_Yc_Y\biggr)^2 \\
=&\sum_Yb_{YX}^2c_Y-\biggl(\sum_Yb_{YX}c_Y\biggr)^2.
\end{split}
\end{equation}

Expressions for $\langle V\rangle_X$, $\Var(V)_X$, $\langle \Delta E^{\text{B}}\rangle_X$ and $\Var(\Delta E^{\text{B}})_X$ follow
trivially from substitution of Eqns. \eqref{QXR_mean} and \eqref{QXR_Var} into Eqns. \eqref{VX_mean}, \eqref{VX_Var}, \eqref{VtotX_mean}
and \eqref{VtotX_Var} respectively:
\begin{equation}
\langle V\rangle_X=a(1-\Lambda)(b_X-\langle b\rangle),
\end{equation}
\begin{equation}
\Var(V)_X=a^2\Lambda^2\omega \Var(b),
\end{equation}
\begin{equation}
\langle \Delta E^{\text{B}}\rangle_X
=\Biggl[\biggl(\frac{1}{r_X^{\text{eff}}}-a\biggr)\Lambda+a\Biggr](b_X-\langle b\rangle) + \Theta_X,
\end{equation}
\begin{equation}\label{VtotXR_Var}
\Var(\Delta E^{\text{B}})_X=\biggl(\frac{1}{r_X^{\text{eff}}}-a\biggr)^2\Lambda^2\omega \Var(b).
\end{equation}

Expressions for $\tilde{E}_L$, $\tilde{E}_M$ and $\tilde{E}$ for random alloys can be obtained by
substituting Eqns. \eqref{QXR_mean} and \eqref{QXR_Var} into Eqns. \eqref{EL}, \eqref{EM} and \eqref{E}, and 
then simplifying:
\begin{equation}\label{ELR}
\tilde{E}_L=\frac{1}{2}a\bigl[\Lambda^2(\omega+1) -2\Lambda +1\bigr]\Var(b) + \frac{1}{2}a\langle b\rangle^2
\end{equation}
\begin{equation}\label{EMR}
\tilde{E}_M=\frac{1}{2}a\bigl[-\Lambda^2(\omega+1)+\Lambda\bigr]\Var(b)
\end{equation}
and
\begin{equation}\label{ER}
\tilde{E}=\frac{1}{2}a(1-\Lambda)\Var(b) + \frac{1}{2}a\langle b\rangle^2.
\end{equation}
In deriving these equations we have used Eqn. \eqref{sumcY=1} and the fact that 
\begin{equation}
\sum_Xb_X^2c_X=\Var(b)+\langle b\rangle^2.
\end{equation}

\subsection{Ternary random alloys}
There have been many experimental \cite{Cole_1997_PRL,Cole_1998,Lewis_1999,Newton_2000,Newton_2004,Marten_2009} and theoretical
\cite{Faulkner_1998_PRL,Marten_2005,Marten_2009} investigations of how $\Var(\Delta E^{\text{B}})_X$ varies with composition in binary random alloys -
a review of which is given in Ref. \onlinecite{Cole_2010}. There have also been many investigations into how $E_M$ varies with composition in binary
random alloys \cite{Magri_1990,Wolverton_1995,Wolverton_1996,Korzhavyi_1994}. However, as far as the authors are aware there have been no such studies
for ternary random alloys.
We will now use expressions derived in the previous subsection to investigate how $\Var(Q)_X$, $\Var(\Delta E^{\text{B}})_X$ and $\tilde{E}_M$ depend on
composition in ternary random alloys constructed from three generic species - which we denote as $A$, $B$ and $C$. To do this we will assume that
the following quantities are composition-independent: the underlying lattice, $a$, the `electropositivity differences' $b_{YX}$ for all $Y$ and $X$,%
\footnote{As alluded to in Sec. \ref{sec:numerical}, one can regard $b_{YX}$ as the electropositivity difference between species $Y$ and
$X$ within the system under consideration so long as $aR_{\text{WS}}$ is sufficiently high. While this is not strictly valid for lower values of 
$aR_{\text{WS}}$, in random alloys one can still regard $b_{YX}$ as the electropositivity difference between species $Y$ and $X$, not with regards to
\emph{every} charge transfer between an $X$ and $Y$ site, but \emph{on average}. This can be seen from the equation
$\langle Q\rangle_Y-\langle Q\rangle_X=\Lambda b_{YX}$, which holds in random alloys and can be derived from Eqn. \eqref{QXR_mean}, and noting that 
$\Lambda>0$ (see Fig. \ref{fig:aG0_vs_aRWS}).}
and the values of $r_X^{\text{eff}}$ for all $X$. We will use the convention that species $A$ is the most electronegative in the alloy, species $C$ is
the most electropositive, and species $B$ has an electronegativity/electropositivity between those of species $A$ and $C$. In other words,
$b_{CA}\geq 0$ and $b_{BA}\geq 0$, where $b_{CA}\geq b_{BA}$. Fig. \ref{fig:CEFM_parameters} provides justification for the assumption that $a$ and
the values of $b_{YX}$ are composition-independent. As can be seen from the figure, the variation in $a_{\,\text{Cu}}$, $a_{\,\text{Zn/Pd}}$ and
$(b_{\,\text{Cu}}-b_{\,\text{Zn/Pd}})$ with $c_{\text{Cu}}$ is, at most, $\approx 10$\%, which implies that the forthcoming results will be at least 
qualitatively accurate. The results of the next section provide some justification for the assumption that $r_X^{\text{eff}}$ is composition-independent.

\begin{figure}
\centering
\subfigure{
\includegraphics[height=0.4\textwidth,angle=270]{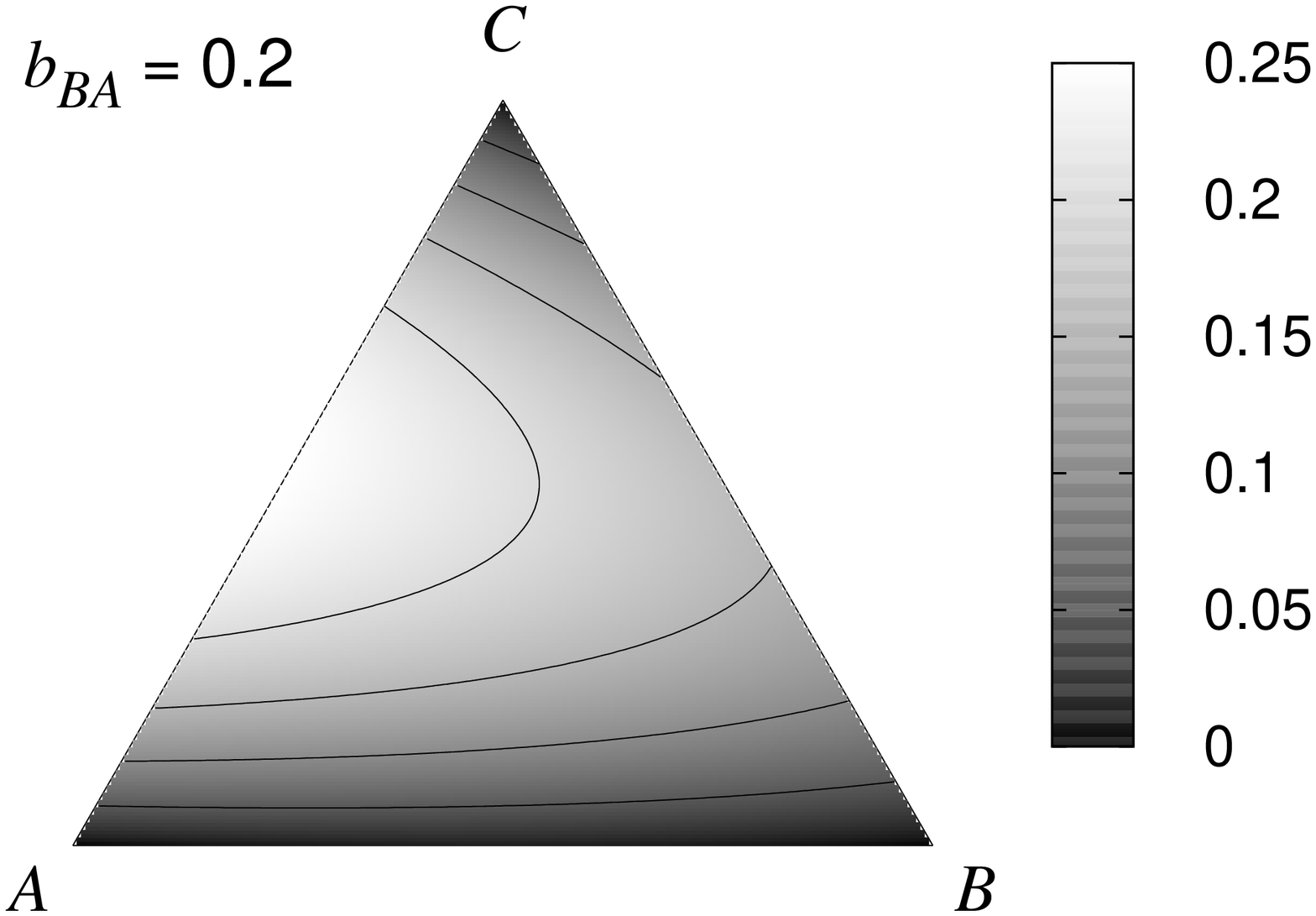}
}
\subfigure{
\includegraphics[height=0.4\textwidth,angle=270]{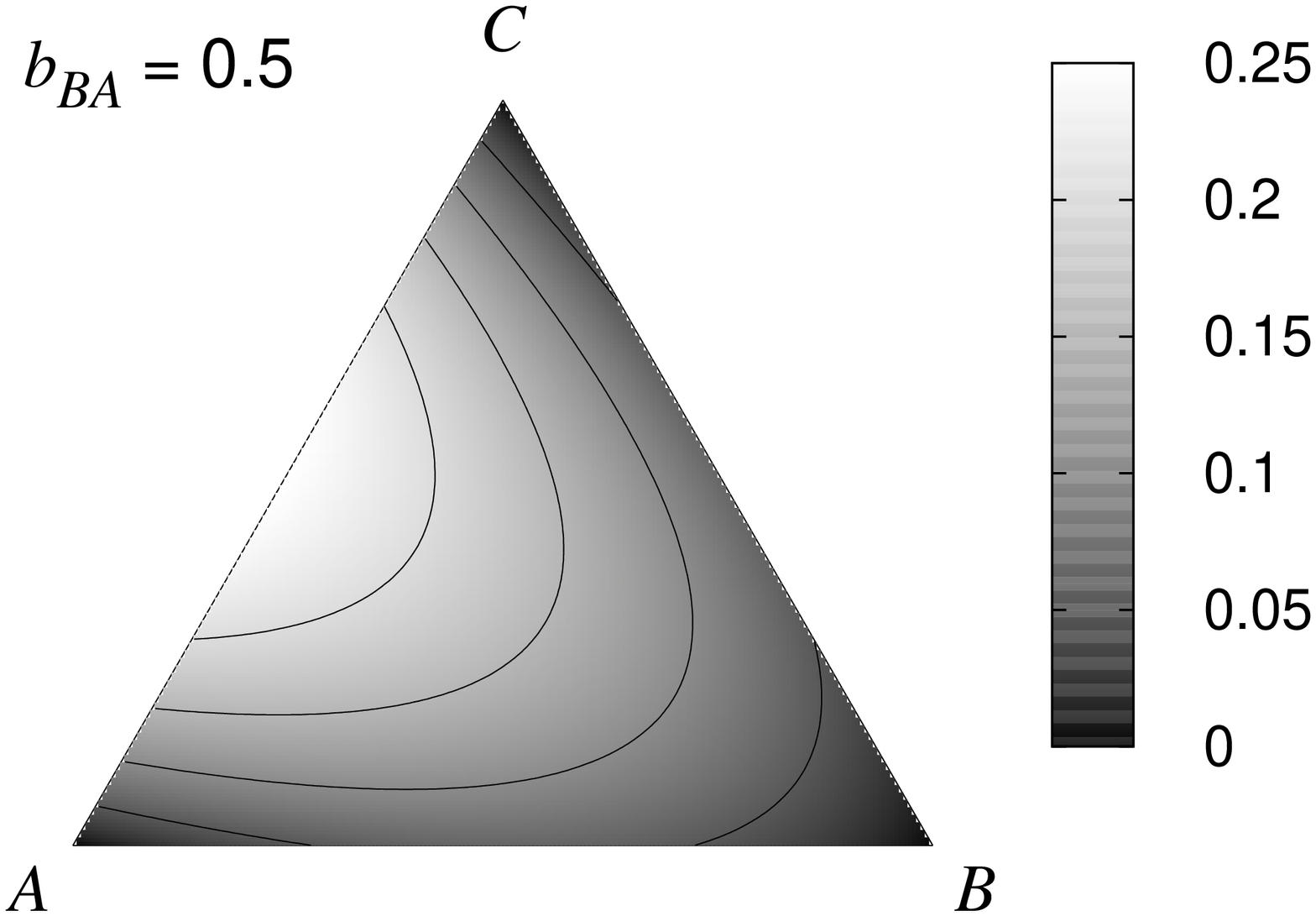}
}
\caption{Ternary graphs of $\Var(b)$ for $b_{BA}=0.2$ and $b_{BA}=0.5$ given that $b_{CA}=1$. The corresponding value of $b_{BA}$ for each graph is
indicated. In each graph, the contours correspond to curves along which $\Var(b)$ is constant. Furthermore, the `outermost' and
`innermost' visible contours in each graph correspond to $\Var(b)=0.05$ and 0.2 respectively. The change in $\Var(b)$ between adjacent
contours is 0.05.}
\label{fig:j}
\end{figure}

We begin by deriving an expression for $\Var(b)$. Setting $X=A$ in Eqn. \eqref{Var_b}, and then simplifying the resulting equation gives
\begin{equation}\label{j}
\Var(b)=b_{BA}^2c_B(1-c_B)+b_{CA}^2c_C(1-c_C)-2b_{BA}b_{CA}c_Bc_C.
\end{equation}
Ternary graphs of $\Var(b)$ are shown in Fig. \ref{fig:j}. The figure contains two graphs. In each graph $b_{CA}=1$ and either:
$b_{BA}=0.2$, which corresponds to species $B$ being more similar, in terms of its electropositivity, to species $A$ than species $C$; or
$b_{BA}=0.5$, which corresponds to species $B$ having an electropositivity exactly halfway between those of species $A$ and $C$. The analogous plot
for $b_{BA}=0.8$, which corresponds to species $B$ being more similar to species $C$ than species $A$, can be obtained from the plot for $b_{BA}=0.2$
by reflecting the $\Var(b)$ surface about the $c_A=c_C$ line.
Fig.  \ref{fig:j} illustrates the following properties of $\Var(b)$, which can be verified analytically using Eqn. \eqref{j}. Firstly,
$\Var(b)$ takes its minimum value of 0 when either $c_A=1$, $c_B=1$ or $c_C=1$. Secondly, it takes its maximum value of $b_{CA}^2/4$ when 
$c_A=c_C=0.5$ - which is the composition at which the two species in the alloy with the largest electropositivity difference are equal,
while there is a vanishing concentration of the species with the intermediate electropositivity. Now, as can be seen from Eqns. \eqref{QXR_Var},
\eqref{VtotXR_Var} and \eqref{EMR}, $\Var(Q)_X$, $\Var(\Delta E^{\text{B}})_X$ and $\tilde{E}_M$ are all proportional to $\Var(b)$, with
composition-independent proportionally constants - as follows from our above assumptions. The proportionality constants can be seen to be positive 
for $\Var(Q)_X$ and $\Var(\Delta E^{\text{B}})_X$%
\footnote{Inspection of Eqn. \eqref{omega_def} reveals that $\omega>0$.}
and so the variation in these quantities with composition will mimic that of $\Var(b)$. For $\tilde{E}_M$ the proportionality constant is instead
negative,%
\footnote{Fig. \ref{fig:aG0_vs_aRWS} and the discussion in Sec. \ref{sec:screening} reveals that $\Lambda\geq 1$. This, in conjunction
with the fact that $\omega>0$, implies that the proportionality constant for $\tilde{E}_M>0$ for $a>0$ - which we have tacitly assumed throughout this
paper.}
and so $\tilde{E}_M$ varies with composition in the opposite sense to $\Var(b)$: $\tilde{E}_M$ is 0 at $c_A=1$, $c_B=1$ or $c_C=1$; and is
\emph{minimized} when $c_A=c_C=0.5$. With regards to  $\Var(\Delta E^{\text{B}})_X$, the NRA-CEFM, in conjunction with the assumptions described at the 
beginning of this section and in Sec. \ref{subsec:CLSs}, therefore predicts that the initial state disorder broadening in ternary random alloys is
maximized at the composition $c_A=c_C=0.5$. Interestingly, this composition is not that with the highest entropy: for a random alloy containing $S$
species, the entropy is maximized at the composition%
\footnote{For $n$ sites which can each belong to $S$ species, the number of possible arrangements $\Omega$ of the system which contain $n_X$
$X$ sites is given by the multinomial coefficient: $\Omega=n!/(\prod_Xn_X!)$. Applying Stirling's approximation, and noting that $n=\sum_Xn_X$ and 
$c_X=n_X/n$, it can be shown that $\ln\Omega=-n\sum_Xc_X\ln c_X$. Since $\ln \Omega$ increases monotonically with $\Omega$, $\Omega$ is maximized when 
$\ln\Omega$ is maximized. Using the method of Lagrange multipliers to impose the constraint described by Eqn. \eqref{sumcY=1}, it can be shown that
$\ln\Omega$ is maximized when Eqn. \eqref{max_entropy_comp} applies.}
\begin{equation}\label{max_entropy_comp}
c_X=1/S\quad\text{for all $X$},
\end{equation}
which for $S=3$ gives $c_A=c_B=c_C=1/3$. By contrast, for \emph{binary} random alloys, the initial state disorder broadening \emph{is} maximized at the
composition with the highest entropy. This can be seen by setting $c_C=0$ in Eqn. \eqref{j} to retrieve the analogous expression for binary random alloys:
\begin{equation}
\Var(b)=b_{BA}^2c_B(1-c_B).
\end{equation}
The above expression is maximized at $c_A=c_B=0.5$, which is the composition with the highest entropy - as can be seen by setting
$S=2$ in Eqn. \eqref{max_entropy_comp}. Hence $\tilde{E}_M$ is minimized, and $\Var(\Delta E^{\text{B}})_X$ for $X=A,B$ is maximized, at $c_A=c_B=0.5$ for
binary random alloys. In fact, the same predictions pertaining to binary random alloys were made using the LCM in the 1990s \cite{Magri_1990,Cole_1998}.

\section{Disorder broadening in CuZn and CuPd random alloys}\label{sec:ab_initio}
We will now apply the NRA-CEFM to real systems.
In Ref. \onlinecite{Faulkner_1998_PRL} the distribution of $\Delta E^{\text{B}}_i$ was determined using the LSMS method for the following random
alloys: fcc Cu$_{50}$Pd$_{50}$, fcc Cu$_{80}$Pd$_{20}$, bcc Cu$_{50}$Zn$_{50}$, and fcc Ag$_{50}$Pd$_{50}$. For the first three of these systems,
plots of $\Delta E^{\text{B}}_i$ vs. $(2Q_i/R_1+V_i)$ for each species are well described by straight lines. This trend can be 
accounted for by the ESCA potential model with a species-dependent $r_i^{\text{eff}}$ as described in Sec. \ref{subsec:CLSs}
\cite{Faulkner_1998_PRL_comment,Faulkner_1998_PRL_reply,Underwood_2009}. With this in mind, we have applied the NRA-CEFM and the ESCA potential model
to the CuPd and CuZn LSMS results, and deduced $r_X^{\text{eff}}$ for each species. We did not consider the AgPd for reasons which will be given at the end of this section.
As can be seen from Fig. \ref{fig:CEFM_parameters}, the NRA holds to a high degree of accuracy in the CuPd and CuZn systems, and hence we expect that the
NRA-CEFM can provide a quantitatively accurate description of the values of $Q_i$ in these systems. We obtained $r^{\text{eff}}_X$ by substituting the LSMS
values of $\Var(\Delta E^{\text{B}})_X$, $a$, $\Var(b)$, and the corresponding values of $\Lambda$ and $\omega$, into Eqn. \eqref{VtotX_Var}, 
and then solving the resulting equation for $r^{\text{eff}}_X$. The LSMS values of $a$ and $\Var(b)$ were calculated from the corresponding values of $a_A$, 
$a_B$ and $b_{BA}$. The specific values of $a_A$, $a_B$ and $b_{BA}$ which were used for each system are as follows.
For the bcc Cu$_{50}$Zn$_{50}$ system we used the `exact' values of $a_{\text{Cu}}$, $a_{\text{Zn}}$ and $b_{\text{CuZn}}$ - which have been determined in Ref. 
\onlinecite{Faulkner_1997}.%
\footnote{The values of $a_X$ and $b_{XY}$ for various CuPd and CuZn random alloys corresponding to the LSMS results of Ref. \onlinecite{Faulkner_1997} are
either tabulated in Ref. \onlinecite{Bruno_2002}, or can be deduced from the quantities tabulated in Ref. \onlinecite{Bruno_2002}.}
The accuracy of the resulting description of this system provided by the NRA-CEFM is illustrated by the fact that Eqn. \eqref{EMR} yields 
-2.47 mRy for $\tilde{E}_M$, which is in excellent agreement with the LSMS values of -2.46 mRy and -2.51 mRy (which correspond to two different random alloy supercells)
quoted in Ref. \onlinecite{Faulkner_1997}. We emphasize that the NRA-CEFM result was obtained relatively easily, i.e., using an analytical formula, whose variables 
are obtained from the numerically calculated $\Lambda$ and $\omega$ at the appropriate value of $aR_{\text{WS}}$ - as tabulated in Ref. 
\onlinecite{supp_material}.
For the CuPd systems the exact values of $a$ and $b_{\text{CuPd}}$ have not been determined. For Cu$_{50}$Pd$_{50}$ we instead used the $aR_{\text{WS}}$ and 
$b_{\text{CuPd}}$ corresponding to a different lattice parameter (7.1 bohrs instead of 6.9 bohrs) which have been determined in Ref. 
\onlinecite{Faulkner_1997}. While it is expected that $aR_{\text{WS}}$ and $b_{BA}$ will vary with lattice parameter,
there are tentative suggestions that for small changes in the lattice parameter these quantities change only very little \cite{Wolverton_1996}. We therefore
expect our results for Cu$_{50}$Pd$_{50}$ to be at least semiquantitatively accurate, and that the general conclusions we draw later are unaffected by the
loss of accuracy incurred by not using the `exact' values of $aR_{\text{WS}}$ and $b_{\text{CuPd}}$ for this system.
For Cu$_{80}$Pd$_{20}$ we used the LSMS values of $a$ and $b_{\text{CuZn}}$ for Cu$_{75}$Pd$_{25}$ (with the `correct' lattice parameter) \cite{Faulkner_1997}. 
As can be seen from Fig. \ref{fig:CEFM_parameters}, the variation in these quantities with $c_{\text{Cu}}$ is very small near $c_{\text{Cu}}=0.75$. Hence our 
choice of $a$ and $b_{\text{CuPd}}$ for Cu$_{80}$Pd$_{20}$ can be considered to be exact for all intents and purposes.

\begin{table*}
\begin{ruledtabular}
\begin{tabular}{cc c cc cc}
\multirow{2}{*}{System} & \multirow{2}{*}{$X$} & LSMS  & \multicolumn{4}{c}{NRA-CEFM} \\
                        &                      & $m_X$ & $R_1/r_X^{\text{eff}}$ & $m_X$     & $R_1/r_X^{\text{eff}}$ & $m_X$     \\
\hline
fcc Cu$_{50}$Pd$_{50}$ & Cu (2p) & -0.01(4) & 2.9(2)  & -0.1(2)  & 2.8(1) & 0.1(2) \\ 
                      & Pd (3d) & -0.13(4) & 3.0(1)  & -0.1(1)  & -      & -      \\
fcc Cu$_{80}$Pd$_{20}$ & Cu (2p) & 0.00(4)  & 3.1(2)  & -0.1(2)  & 2.9(2) & 0.1(1) \\ 
                      & Pd (3d) & -0.11(4) & 3.1(1)  & -0.1(1)  & -      & -      \\ 
bcc Cu$_{50}$Zn$_{50}$ & Cu (2p) & 0.48(4)  & 3.31(4) & 0.44(2)  & -      & -      \\
                      & Zn (2p) & 0.62(4)  & 2.83(2) & 0.65(1)  & -      & -      \\
\end{tabular}
\end{ruledtabular}
\caption
{$R_1/r_X^{\text{eff}}$ and $m_X$ calculated using the combined NRA-CEFM and ESCA potential model for various random alloys, and the analogous values
of $m_X$ obtained from LSMS calculations \cite{Faulkner_1998_PRL}. $m_X$ is the gradient of the curve $\Delta E^{\text{B}}_i$ vs. $(2Q_i/R_1+V_i)$. Uncertainties
in the model values stem from uncertainties in the LSMS values from which they are ultimately derived; and the aforementioned uncertainties in the LSMS values 
reflect the finite precision to which they are quoted in Ref. \onlinecite{Bruno_2002}. The LSMS values of $m_X$ were obtained from Fig. 1 in Ref. 
\onlinecite{Faulkner_1998_PRL}, and the uncertainties on these quantities are approximate.}
\label{table:r_eff}
\end{table*}

The $R_1/r^{\text{eff}}_X$ obtained by solving Eqn. \eqref{VtotX_Var} are presented in Table \ref{table:r_eff}. Note that Eqn. \eqref{VtotX_Var} has two possible 
solutions for $r_X^{\text{eff}}$. The `correct' solution can be deduced by comparing the predicted gradient $m_X$ of the curve $\Delta E^{\text{B}}_i$ vs. 
$(2Q_i/R_1+V_i)$ for each solution to that of the analogous LSMS curve. Unfortunately the correct solution cannot not be unequivocally deduced for Cu in the
CuPd systems. For these systems we give both solutions for $R_1/r^{\text{eff}}_X$. However, this is inconsequential because the two solutions are so similar as 
not to affect our forthcoming conclusions.
In Ref. \onlinecite{Underwood_2009} it was found that OLCM simulations utilizing $R_1/r_X^{\text{eff}}=3$ gives good agreement with the LSMS value of 
$\Var(\Delta E^{\text{B}})_X$ for Cu$_{50}$Zn$_{50}$, leading the authors to hypothesize that $R_1/r_X^{\text{eff}}=3$ should be used instead of 
$R_1/r_X^{\text{eff}}=2$ - as had been used previously - in the ESCA potential model when applied to disordered alloys. Our results strongly support this hypothesis,
though they do reveal that there are small variations in $R_1/r_X^{\text{eff}}$ between species within the same system and between different systems.

\begin{table*}
\begin{ruledtabular}
\begin{tabular}{cc c c cc cc}
System & $X$ & FWHM & $a$ & $1/r_X^{\text{eff}}$  & $a_X^{\text{tot}}$ & $1/r_X^{\text{eff}}$ & $a_X^{\text{tot}}$ \\
\hline
fcc Cu$_{50}$Pd$_{50}$ & Cu (2p) & 0.05 & 16.06(1) & 16.4(9) & 0.4(9)  & 15.7(8) & -0.4(8) \\ 
                      & Pd (3d) & 0.10 & 16.06(1) & 16.8(6) & 0.7(6)  & -       & -       \\
fcc Cu$_{80}$Pd$_{20}$ & Cu (2p) & 0.05 & 16.16(2) & 17(1)   & 0(1)    & 15.7(9) & -0.5(9) \\ 
                      & Pd (3d) & 0.08 & 16.16(2) & 16.9(8) & 0.7(8)  & -       & -       \\ 
bcc Cu$_{50}$Zn$_{50}$ & Cu (2p) & 0.35 & 24.87(3) & 18.9(2) & -6.0(3) & -       & -       \\
                      & Zn (2p) & 0.51 & 24.87(3) & 16.2(1) & -8.7(1) & -       & -       \\
\end{tabular}
\end{ruledtabular}
\caption
{Contributions to the disorder broadening in various random alloys. The FWHM and $a$ pertain to the LSMS results of Ref. \onlinecite{Faulkner_1998_PRL}, while the 
remaining quantities were determined from these using the combined NRA-CEFM and ESCA potential model. The FWHM values are in eV, while the remaining quantities are 
in volts. Uncertainties in the model values stem from uncertainties in the LSMS values from which they are ultimately derived; and the aforementioned uncertainties
in the LSMS values reflect the finite precision to which they are quoted in Ref. \onlinecite{Bruno_2002}.}
\label{table:broadening_breakdown}
\end{table*}

The LSMS magnitudes of disorder broadening for the CuPd and CuZn systems are given in Table \ref{table:broadening_breakdown}. Note that the broadening is
significantly larger in CuZn than CuPd. Our results provide insight as to why this is the case.
Consider Eqn. \eqref{VtotX}, which gives $\Delta E_i^{\text{B}}$ for an $X$ site within the combined NRA-CEFM and ESCA potential model. Note that
$\Delta E_i^{\text{B}}$ is linear in $Q_i$, with proportionality constant 
\begin{equation}
a^{\text{tot}}_X\equiv 1/r^{\text{eff}}_X-a.
\end{equation}
This reflects the fact that shifts in the \emph{total} electrostatic potential at $X$ nuclei are the sum of two contributions which are both linear in $Q_i$:
the shift in the \emph{intra-site} electrostatic potential, which has proportionality constant $1/r^{\text{eff}}_X$; and the shift in the Madelung potential,
which has proportionality constant $-a$ (see also Eqns. \eqref{Vtot_ESCA} and \eqref{QV_rel}). Hence $a^{\text{tot}}_X$, which determines the magnitude of the 
disorder broadening (see Eqn. \eqref{VtotXR_Var}), can be regarded as a combination of intra-site and Madelung broadenings. Note that these broadenings cancel
due to the fact that their respective proportionality constants have opposite signs ($a>0$). In Table \ref{table:broadening_breakdown} the values of 
$1/r^{\text{eff}}_X$ and $a$ - which determine the magnitudes of the intra-site and Madelung broadenings respectively - are compared with $a^{\text{tot}}_X$ 
for the CuPd and CuZn systems. As can be seen from the table, the intra-site broadening is approximately the same magnitude for all species in all systems;
it is the fact that the Madelung broadening is significantly larger in CuZn which leads to the significantly larger disorder broadening in CuZn compared to
CuPd. Interestingly, the Madelung and intra-site broadenings are very similar in magnitude for CuPd, the result being that they almost exactly cancel,
leaving a very small `total' broadening. Furthermore, $a_X^{\text{tot}}$ is positive for Pd in Cu$_{50}$Pd$_{50}$, which reflects the fact that the intra-site 
broadening is larger than the Madelung broadening, while $a_X^{\text{tot}}$ is negative for Cu and Zn in CuZn, which reflects the fact that the Madelung 
broadening dominates. With regards to Eqn. \eqref{VtotX}, the aforementioned signs of $a_X^{\text{tot}}$ mean that $\Delta E_i^{\text{B}}$ varies with $Q_i$ for
Pd in CuPd in the opposite sense to Cu and Zn in CuZn.

To conclude this section we will briefly comment on fcc Ag$_{50}$Pd$_{50}$ - the remaining random alloy considered in Ref. \onlinecite{Faulkner_1998_PRL}. For
this system, as opposed to CuPd and CuZn considered above, the LSMS plot of $\Delta E^{\text{B}}_i$ vs. $(2Q_i/R_1+V_i)$ for each species cannot be
well described as linear. This implies in Ag$_{50}$Pd$_{50}$ either: that the assumption that $r_i^{\text{eff}}$ takes the same value for all $X$ sites
is invalid; that the whole ESCA potential model itself is invalid; or that the \emph{Q-V} relations do not hold. The first of these possibilities
is most likely, though further investigation is required. Differences in the values of $r_i^{\text{eff}}$ of $X$ sites within a given system could possibly be 
accounted for by decomposing the valence charge within each site into components corresponding to each angular momentum quantum number $l$, and assuming that 
the effective radius corresponding to each $l$-component takes the same value for all $X$ sites. This approach has been applied to the \emph{average} of $X$ 
core level shifts in alloys - as summarised in Ref. \onlinecite{Cole_2010}. However, there have been no attempts to use this approach in order to determine 
the \emph{distribution} of $X$ core level shifts with a given alloy.

\section{Summary}\label{sec:summary}
We conclude this paper by giving a summary of our key findings.
We began by deriving the CEFM energy function in order to elucidate its underlying approximations. These approximations
are: the spherical approximation; that the site charges are perturbed from their `bare' values by only a small amount; and that $E_{L,i}$
- the `non-Madelung' contribution to the total energy from site $i$ - is a functional only of the contents of site $i$, and not of the contents of
any other site. Three ways in which the last of these approximations can be achieved were highlighted: if outwith site $i$ is assumed to be an effective
medium in the evaluation of $E_{L,i}$; if both the Thomas-Fermi and local density approximations are utilized; and if `atomic boundary conditions' are used
in the evaluation of $E_{L,i}$.

We then limited our scope to the particular case of the CEFM in which the strength of the `local interactions' within each site are the same for all sites.
The properties of this model - the NRA-CEFM - were explored in detail. In Section \ref{sec:fundamental_properties} it was shown that the net charges in 
the NRA-CEFM can be understood as resulting from charge transfer between all pairs of sites in the same manner as the optimised linear charge model for the 
case of binary alloys, and hence can be considered to be the generalization of the optimised linear charge model for alloys containing any number of chemical 
species.

In Section \ref{sec:screening} the `geometric factors' in the NRA-CEFM were determined for fcc, bcc and sc lattices, and the nature of the
screening in the model was explored. An analytical description of the screening was deduced for the limit of weak inter-site Coulomb interactions.
Here, the nature of the screening was shown to be universal, i.e. the same for all systems. Numerical calculations were used to determine
the nature of the screening away from this limit. The results of these calculations were used to illustrate how the universality in the screening
increasingly breaks down as the strength of the inter-site Coulomb interactions is increased. At the end of Section \ref{sec:screening} our results were 
compared to those of Ref. \onlinecite{Ruban_2002}, and found to be in quantitative agreement. This was attributed to the fact that all of the approximations 
which underpin the NRA-CEFM are either implicit in the electronic structure calculations of Ref. \onlinecite{Ruban_2002} or can be justified 
\emph{a posteriori} from the results of that study.

In Section \ref{sec:analytical} we used the NRA-CEFM to derive analytical expressions for various physical quantities which can be applied
to any system. These physical quantities include the mean and variance in the charges for each species, the initial state core level shifts, and the 
Madelung and total energies of the alloy. Analogous expressions were then derived for random alloys. These expressions were then used to investigate how the variance 
in the net charges for each species, the magnitude of the core level initial state disorder broadening, and the Madelung energy, depend on composition
in ternary random alloys. The magnitudes of these quantities were shown to be maximized at the composition where the two species in the alloy with the
largest electronegativity difference have equal concentrations, and the remaining species has a vanishing concentration. With regards to the disorder broadening,
as opposed to the case for binary random alloys, in ternary alloys this composition does not correspond to that with the highest entropy.

In Section \ref{sec:ab_initio} we applied the NRA-CEFM to CuPd and CuZn random alloys. The model was used to determine the effective radius associated
with valence electron charge transfer in the ESCA potential model for these systems. These radii were found to be $R_1/3$, where $R_1$ is the
nearest neighbor distance, with only a small species- and system-dependence. Our results were then used to examine how the separate disorder broadenings
associated with the intra-site electrostatic and Madelung potentials contribute to the `total' disorder broadening. In CuZn it was found that the Madelung broadening
dominates, while for Pd in CuPd the intra-site broadening dominates. The result is that a site's core level shift depends on its net charge - which characterizes the 
site's environment - in the opposite sense for Pd in CuPd than for Cu or Zn in CuZn.

We expect that our analytical and numerical results will enable the models studied here to provide a simple yet accurate framework for the interpretation
of XPS and Auger electron spectroscopy core level disorder broadening, as well as for the investigation of segregation and ordering phenomena, in 
a wide range of alloy systems.

\begin{acknowledgements}
Valuable discussions with Graeme Ackland are greatly acknowledged. This work was supported by the Engineering and Physical Sciences Research Council.
\end{acknowledgements}


\end{document}